\RequirePackage{fix-cm}
\documentclass[twocolumn,epjc3]{svjour3}  
\smartqed  
\RequirePackage{graphicx}
\journalname{Eur. Phys. J. C}
\usepackage{amssymb}
\usepackage{amsmath}

\def\be{\begin{eqnarray} &&}
\def\nonu{\nonumber \\ &&}
\def\ee{\end{eqnarray}}
\def\beq{\begin{equation}}
\def\eeq{\end{equation}}
\newcommand{\mbf}[1]{\mathbf{#1}}
\usepackage{color}

\begin{document}

\title{Solving  the inhomogeneous Bethe-Salpeter Equation 
in  Minkowski space: the zero-energy limit}


\author{Tobias Frederico \thanksref{e1,addr1}\and Giovanni Salm\`e
\thanksref{e2,addr2}\and
Michele Viviani \thanksref{e3,addr3}}

\thankstext{e1}{e-mail: tobias@ita.br}
\thankstext{e2}{e-mail: salmeg@roma1.infn.it}
\thankstext{e3}{e-mail: michele.viviani@pi.infn.it}

\institute{Dep. de F\'\i sica, Instituto Tecnol\'ogico de Aeron\'autica,
DCTA, 12.228-900 S\~ao Jos\'e dos
Campos, S\~ao Paulo, Brazil \label{addr1}\and Istituto  Nazionale di Fisica Nucleare, Sezione di Roma, P.le A. Moro 2,
 I-00185 Roma, Italy \label{addr2}\and Istituto  Nazionale di Fisica Nucleare, Sezione di Pisa,
Largo Pontecorvo 3, 56100, Pisa, Italy \label{addr3} 
}

\date{Received: date / Accepted: date}

\maketitle

\begin{abstract}
For the first time, the inhomogeneous Bethe-Salpeter Equation for an interacting
  system, composed by two massive scalars
exchanging a massive scalar, is numerically investigated in ladder
approximation,  directly in Minkowski space, by using an approach based on the Nakanishi integral
representation. In this paper,    the limiting 
case of zero-energy states is considered, 
 extending the approach successfully applied to  bound 
states. The  numerical values of scattering lengths, {are calculated for several values of the Yukawa coupling constant},
 by using two different integral
equations that stem within the Nakanishi framework. { Those low-energy observables  } are compared with (i) the
 analogous
quantities recently obtained in { literature},
within a totally different framework and {(ii) the  non relativistic
evaluations, for  illustrating the  relevance of a non perturbative, genuine field
theoretical  treatment in Minkowski
space, even in the  
low-energy regime. }
 Moreover, { dynamical functions, like}  the Nakanishi weight functions and
 the distorted part of the zero-energy
Light-front wave functions are also presented.
Interestingly, a highly non trivial issue related to the abrupt change  in the
 width of 
the support of the Nakanishi weight function, when the zero-energy limit is
approached, 
is elucidated, { ensuring a sound basis to }  the forthcoming  evaluation of phase-shifts.

\end{abstract}
\section{Introduction}
\label{Intr}
{ Within a  field 
theoretical
framework, it is a highly non trivial challenge to develop non perturbative tools in Minkowski space, but it is quite desirable
to devote efforts  in  that direction,  in order to gain insights that could turn out useful  in particle physics. }
In the last few years,  solving the homogeneous Bethe-Salpeter
equation (BSE) \cite{SB_PR84_51}, directly in Minkowski space, has made a substantial step
forward
\cite{KWa,KWb,carbonell1,carbonell2,carbonell3,carbonell4,carbonell5,FSV1,FSV2,FSV3} due to approaches based on the so-called  Nakanishi
perturbation-theory integral representation (PTIR) of the $n$-leg transition
amplitudes \cite{nak71}.

 The Nakanishi PTIR  for the three-leg 
 amplitude is emerging as a very  effective tool for studying the bound state
 problem \cite{KWa,KWb,carbonell1,carbonell2,carbonell3,carbonell4,carbonell5,FSV2,FSV3}, within a rigorous field-theory framework. Though
the Nakanishi PTIR of the three-leg  amplitude, or vertex
function, had been devised within the {\em perturbative} framework of the Feynman
diagrams (as  it happens for any  $n$-leg amplitude PTIR),  it  has been shown to
work
  extremely well as { the initial}  Ansatz for obtaining  {\em actual 
  solutions} of  the homogeneous BSE. 
  It must be recalled that BSE, being an integral
  equation, belongs to a  non perturbative realm, and therefore the Nakanishi integral representation of the  three-leg
  amplitude can be only an Ansatz, when exploited in this context.
  
The main features of the Nakanishi integral representation
of any $n$-leg amplitude are basically related 
to the formal  infinite sum of the parametric
 Feynman 
diagrams, that contribute to the amplitude under
consideration.
In particular, the $n$-leg amplitude PTIR has a well-defined structure, given
by
 the folding of (i) a denominator,  containing all
the allowed
independent invariants and  governing  the analytic behavior of the amplitude
itself, and (ii)
  a 
  weight function, that  is a real function depending upon real
  variables (one is a non compact variable, while  the others are compact). It
  should be emphasized that, at this stage, the Nakanishi weight function has 
   only 
   a formal expression \cite{nak71}.
 If there were an equation for explicitly determining  such a weight function, 
 then one could quantitatively evaluate  the  actual 
  $n$-leg amplitude, under consideration.
   The homogeneous BSE, that {obviously does not belong to }
    the original framework of PTIR,
  has inspired a different usage of the formal expression of a particular $n$-leg
  amplitude, namely the three-leg one, {or vertex function}. Indeed,
   if one assumes that the  Nakanishi integral representation  of 
   the three-leg amplitude  be formally valid also for the BS 
   amplitude (still a three-leg
   amplitude, but for a bound state), then the  weight function could be 
   considered as an unknown function to be determined. { It has to be pointed out that, {\em a priori}, 
   there is no guarantee that such an approach for solving BSE be successful, given the caveat above mentioned. Fortunately, it
   works, as shown in  
    Refs. 
    \cite{KWa,KWb,carbonell1,carbonell2,carbonell3,carbonell4,carbonell5,FSV2,FSV3}, where the above strategy was
    applied, but with some differences,} 
     for solving  the homogeneous BSE directly in Minkowski space.
  More precisely, by using the PTIR Ansatz for the BS amplitude one can
 derive, {  in a {\em formally exact} way},  an equation for the Nakanishi weight function, starting from the
  homogeneous BSE, and look for  solutions. If 
 { the new equation for the weight function has solution}, 
 then one can
 claim that   BS amplitudes, actual solutions of the homogeneous BSE in
  Minkowski space, can be 
  (i)  formally written like the PTIR three-leg amplitude, and (ii) 
    numerically determined. {In order to achieve 
   a  formally 
 exact 
 integral equation for the weight
 function   from BSE, it is  
  very  useful and effective   adopting a  Light-front (LF) 
  framework. This has been done both  in the  covariant version of the LF framework
 \cite{carbonell1} and in the non-explicitly covariant one  \cite{FSV1}. }
 In particular,  the bound states of a massive  two-scalar system interacting 
 through the exchange of a
 massive scalar  have been studied by adopting both ladder 
 \cite{carbonell1,carbonell3,carbonell4,FSV2,FSV3} and cross-ladder
 approximations of the BS kernel \cite{carbonell2}. Notably, the extension to a bound fermionic system have been
also undertaken \cite{carbonell5}.
 { It has to be recalled that  numerical investigations of the 
 homogeneous BSE
  has been  performed  also by considering  the 
  standard 4-dimensional variables \cite{KWa,KWb}.} 
 
The successful achievements for the homogeneous BSE encourage
the extension of  the Nakanishi integral representation to
the study of the inhomogeneous BSE, i.e. the integral equation that determines 
the scattering states. Our aim  is to present { a new application} of
our general approach {\cite{FSV1}, based on the so-called  LF projection of the BS amplitude,
 i.e. the exact
integration on the minus component of the relative four-momentum that appears in  the BS amplitude. After applying
 this formally exact
step to BSE, we  have
numerically investigated
 the 
zero-energy limit of the inhomogeneous  BSE, }for  a massive two-scalar system interacting 
 through the exchange of a
 massive scalar, in ladder approximation. The  calculated scattering
lengths   have been  compared in great detail with the analogous  observables recently obtained \cite{carbonell6,carbonell7} 
within a completely
 different
framework. { We have also compared our results
    with the non relativistic 
scattering lengths, with  the intent to   
yield a possible guidance for lowering the model dependence in the treatment of interacting final states, pertaining to relevant hadronic 
decay modes.  { Indeed,  improving  and widening  our study could contribute 
to achieve  an actual evaluation of  the covariant off-shell T-matrix, that
represents a key ingredient
 for describing, e.g., the heavy meson decay 
amplitudes, like in $D\to K\pi\pi$  processes \cite{MagPRD11,GuiJHEP14}, 
and  could also have an impact in the development of final-state-interaction 
models, needed in the analysis of the CP violation in 
charmless three-body $B$ decays \cite{BedPRD14}.}}
 Moreover, we have properly analyzed the distorted part of the
 zero-energy wave function, putting in evidence the relation between a non smooth
behavior of the Nakanishi weight function and the expected singularities of the
LF 3D wave function, like the one that brings the information relative to the global
propagation of the interacting two-scalar system.
Finally, the  integral equation for the Nakanishi weight
function obtained by applying the so-called uniqueness theorem \cite{nak71}, 
is carefully 
analyzed  for the general case of positive energy. Such an
in-depth
analysis allows us to illustrate a surprising change in the width (from
$(-\infty,\infty)$ to
$[0,\infty)$)  of the support
of the Nakanishi weight function with respect to its non compact variable, when
the zero-energy limit is considered. {Clarifying this feature 
 allows us to put  the forthcoming
 calculation of the phase-shifts on a sound basis, since their calculation
requests a careful analysis from both the theoretical and numerical points of 
view, as it will be illustrated  elsewhere \cite{FSV4}.
By concluding this Introduction, it could be useful to remind that 
developing  genuine non perturbative
descriptions of the scattering processes within the Minkowski space,
 possibly  applying
 formally
exact frameworks,  is an appealing goal, in view of attempts of extracting
  tiny, but fundamental signals once
very accurate experimental data will become available.}

 The paper is organized as follows. In Sec. \ref{NIE}, { we shortly introduce
  both   the definitions and the general formalism, and we thoroughly discuss}  
 the problem of the support of the Nakanishi weight function, given its relevance for the 
  zero-energy limit. 
  Sec. \ref{scatle} illustrates how
 to evaluate the scattering
 length from the Nakanishi weight function, in ladder approximation.
 In Sec. \ref{results}, the numerical studies of the scattering length are presented  
 and compared with the existing calculations
 found in literature; moreover 
  the scattering 3D LF wave function (indeed the distorted part) is analyzed.
   Finally, in Sec. \ref{concl}, the conclusions are drawn.
  
\section{The Nakanishi  Integral Equations for  scattering states}
\label{NIE}
In this Section,   (i) we  quickly recall the general formalism of  
 Ref. \cite{FSV1}, for 
obtaining two integral equations 
that allows one to determine the Nakanishi
weight function needed for  scattering processes, and  (ii) 
we demonstrate
a relevant feature of the weight-function support, that { it turns out to be} very important 
 also for numerically solving
 the inhomogeneous BSE. 
 
In our investigation, we considered an interacting system 
  composed by two massive scalars that exchange a massive scalar. This is
   a generalization of the honorable
Wick-Cutkosky model\cite{GW,Cut} in two respects: 
(i) the interaction takes place through a massive-scalar exchange and
(ii) the  
scattering states 
is our  focus. 
  
\subsection{General formalism}
\label{GF}
For scattering states, the  incoming particles are on
their-own mass-shell and we indicate their  total and relative four-momenta with 
$p$ and  $k_i$, respectively.  {By assuming that 
 the inhomogeneous BS amplitude 
$\Phi^+(k,p,k_i)$  be  expressed in terms of the Nakanishi weight-function
 $g^{(+)}(\gamma',z',z'';\kappa^2,z_i)$, then   one  can write (cf
  Ref. \cite{FSV1})}
    \be 
\Phi^{(+)}(k,p,k_i)= \nonu =(2\pi)^4\delta^{(4)}(k-k_i)-
i~\int_{-1}^1dz'\int_{-1}^1dz''\int_{-\infty}^{\infty}d\gamma'\nonu \times
 \frac{g^{(+)}(\gamma',z',z'';\kappa^2,z_i)}{\left[\gamma'+m^2
-\frac{1}{4}M^2-k^2-p\cdot k\; z''-2 k\cdot k_i~z' -i\epsilon\right]^3}=\nonu=
(2\pi)^4\delta^{(4)}(k-k_i)-
i~\int_{-1}^1dz'\int_{-1}^1dz''\int_{-\infty}^{\infty}
d\gamma'
 ~\times \nonu\frac{g^{(+)}(\gamma',z',z'';\kappa^2,z_i)}{\left[{\cal
 D}_0-i\epsilon\right]^3}\, ,
\label{ptirsc}\ee
where   the total four-momentum  is $ p\equiv\{M,{\bf 0}\}$  and
\be{\cal D}_0=\gamma'+\gamma+\kappa^2
-k^-(k^+ +{M \over 2}z''-{M \over 2}z_i z') \nonu -k^+ {M \over 2}( z'' + z_i z') 
 +
 2 z'cos \varphi \sqrt{\gamma
\gamma_i}~~.\nonumber\ee
 The
power of the 
denominator is the same  one adopted for describing a bound state  (cf  Refs.
  \cite{carbonell1,carbonell2,FSV1,FSV2,FSV3}). Exploiting
a standard formalism introduced in Ref. \cite{carbonell1}, one defines $z_i=-2 k^+_i/M$ and gets  
 $z_i=2 k^-_i/M$,  since the incoming particles are on their-own mass
shell: $(p/2\pm k_i)^2=m^2$. Moreover, one has   $1\geq |z_i|$, 
since the incoming particles have positive
longitudinal momenta, i.e. $p^+/2 \pm k^+_i \geq 0$. In Eq. \eqref{ptirsc}, the following notations have been used:
 (i) $cos \varphi =\widehat {\bf k}_\perp \cdot
\widehat{\bf k}_{i\perp}$, (ii)
$\gamma=|{\bf k}_\perp|^2$ and  $\gamma_i=|{\bf
k}_{i\perp}|^2$, and (iii) $\kappa^2=m^2 -M^2/4$. 
 For the initial state one has
\beq
(p/2\pm k_i)^2=m^2={M^2\over 4} +k^+_ik^-_i -\gamma_i= (1-z^2_i){M^2\over 4}-\gamma_i \, ,
\eeq
 with necessarily $(z_i)^2< 1$. Hence  one gets 
 \be
M^2 = 4~ {(m^2+\gamma_i) \over (1-z^2_i)}
\nonu
\kappa^2=-\gamma_i-z^2_i {M^2\over 4}=k^2_i~\leq~0 \, ,
\label{en1}\ee
{ To complete the generalities, we also give the expression for the 
inhomogeneous BSE, 
without self-energy insertions and vertex 
corrections, in the present stage of our approach,. Then,
 one can write}
\be
\Phi^{(+)}(k,p,k_i)=(2\pi)^4\delta^{(4)}(k-k_i)\nonu + G_0^{(12)}(k,p)
~
\int \frac{d^4k^\prime}{(2\pi)^4}i~{\cal K}(k,k^\prime,p)\Phi^{(+)}(k^\prime,p,k_i) \,,
\label{inbse}\ee
where 
$i~{\cal K}$ is the interaction kernel (where  the  vertex corrections should appear), and $G_0^{(12)}$ is the free two-particle  
Green's function given by
\be\label{G0}
G_0^{(12)}(k,p) =G_0^{(1)}G_0^{(2)}=\nonu=
\frac{i}{(\frac{p}{2}+ k)^2-m^2+i\epsilon}~~~
\frac{i}{(\frac{p}{2}-k)^2-m^2+i\epsilon} \, .
\ee
 
It is worth noting that the bosonic
 symmetry of the BS amplitude, Eq. \eqref{ptirsc}, when $1 \to 2$,
(i.e. $p\to p$, $k\to (-k)$ and $k_i\to (-k_i)$) has to be fulfilled, as in the case of  bound states \cite{FSV2}.
Therefore, 
the Nakanishi weight function must have the following property
\be
g^{(+)}(\gamma',z',z'';\kappa^2,z_i)=g^{(+)}(\gamma',z',-z'';\kappa^2,-z_i) \, .
\label{symp}\ee
Moreover,  { as shown in details in
 in 
 \ref{zz'}, one has 
 \be
 g^{(+)}(\gamma',z'=\pm 1,z'')=g^{(+)}(\gamma',z',z''=\pm 1)=0 \,.
\label{edge} \ee
  {As well-known (see e.g. Refs. \cite{Br_rev,CK_rev,FSV1}),} by projecting the BS amplitude onto the
 null-plane, i.e. integrating  on  $k^-$, one {\em exactly} gets
the 3D
LF scattering wave function $\psi^{(+)}$, that is  proportional   to the 
valence component $
 \psi^{(+)}_{n=2/p}$ appearing in the Fock expansion of a two-scalar state, 
 namely 
 $\psi^{(+)}=\sqrt{2}~
 \psi^{(+)}_{n=2/p}$ (given the normalizations assumed in 
  Refs. \cite{FSV1,FSV2}). The 3D LF scattering wave function   reads 
   \be
\psi^{(+)}\left(z,\gamma,cos\varphi;\kappa^2,z_i\right)=\nonu = p^+ {(1-z^2)\over
4} \int {dk^-\over 2 \pi}~ \Phi^{(+)}(k,p,k_i)= p^+ {(1-z^2)\over
4} \nonu \times~(2\pi)^3\delta^{(3)}(\tilde k-\tilde k_i) + \psi_{dist}\left(z,\gamma,cos\varphi;\kappa^2,z_i\right)
\label{scat3d}
\ee
where  $\tilde k \equiv \{k^+, {\bf k}_\perp\}$ and 
$\psi_{dist}\left(z,\gamma,cos\varphi;\kappa^2,z_i\right)$ is the distorted part of 
the 3D LF scattering wave function, that in the CM frame, where  $p^+=p^-=M/2$ and ${\bf
p}_\perp=0$, reads
\be
\psi_{dist}\left(z,\gamma,cos\varphi;\kappa^2,z_i\right)={(1-z^2)\over  4}
\int_{-1}^1 dz' \nonu \times
\int_{-\infty}^{\infty}d\gamma'\frac{g^{(+)}(\gamma',z',z;\kappa^2,z_i)}
{\left[{\cal D}_1-i\epsilon\right]^2}\, .
\ee
with
\be
{\cal D}_1=\gamma'+\gamma+ z^{ 2} m^2+(1- z^{ 2})\kappa^2 \nonu
 +z' ({M ^2\over 2} z~  z_i  +
 2 cos \varphi\sqrt{\gamma
\gamma_i})~~~.
\nonumber \ee

In what follows, without loss of generality, we choose 
 a head-on scattering process, namely a $z$-axis along the incoming 
three-momenta. In this case    the variable $\gamma_i$ is zero and  therefore the dependence upon
$cos\varphi$ disappears. As a matter of fact, the distorted wave function becomes
\be
\psi_{dist}\left(z,\gamma;\kappa^2,z_i\right)= {(1-z^2)\over  4}
\int_{-1}^1 dz'~
\int_{-\infty}^{\infty}d\gamma'
\nonu \times ~\frac{g^{(+)}(\gamma',z',z;\kappa^2,z_i)}
{[ {\cal D}_2 -i\epsilon]^2} 
\label{scat3db}
\ee
with 
\be
{\cal D}_2=\gamma'+\gamma+ z^{ 2} m^2+(1- z^{ 2})\kappa^2
 +z' {M ^2\over 2} z~  z_i
\ee
and
$$z_i= \pm {2\over M}~\sqrt{-\kappa^2}~~~~~~.$$
{Remarkably, $\psi_{dist}\left(z,\gamma;\kappa^2,z_i\right)$  displays  a cut, originated   by  the free propagation of the two 
constituents, just as in the non
relativistic case. In particular,   the distorted part 
of the
scattering wave function can be rearranged in order to make explicit the free
propagation,
obtaining  (see details in \ref{distwf})}
\be
\psi_{dist}\left(z,\gamma;\kappa^2,z_i\right)
= ~i~ {(1-z^2)\over  4}\nonu \times ~
{1\over \left[\kappa^2(1-z^2)+ m^2z^2+\gamma 
-
i\epsilon\right]}\int_{-1}^1 d\zeta''\int_{-1}^1 d\zeta^\prime
\nonu \times
\int_{-\infty}^\infty d\gamma'' ~ \widetilde{\cal G}^+(\gamma'',\zeta'',\zeta^\prime;
\kappa^2,z_i)~\theta (1 -|\zeta''| -|\zeta'|)
\nonu \times ~
\left[{ (1 +z) \over  (1+\zeta' -\zeta'' z_i)} ~ {\theta(\zeta'-z-\zeta'' z_i )~
\over 
{\cal D}_3( z,\zeta',\zeta'')-i\epsilon}
+\right. \nonu \left. +{(1-z)\over (1-\zeta' +\zeta'' z_i)} ~
{\theta(z+\zeta''z_i-\zeta')~
\over {\cal D}_3( -z,-\zeta',-\zeta'')-i\epsilon  }\right] \, ,
\label{psidist}\ee 
where 
\be
{\cal D}_3( z,\zeta',\zeta'')=\kappa^2(1-z^2)+ m^2z^2+\gamma   \nonu + 
 { (1 +z) \over  (1+\zeta' -\zeta'' z_i)} ~\left(
  {M^2\over 2} z\zeta'' z_i
+\gamma''  \right)
\ee
and $\widetilde{\cal
G}^+(\gamma',\zeta'',\zeta^\prime;\kappa^2,z_i)$ is the Nakanishi weight function for the
half-off-shell T-matrix (see Ref. \cite{FSV1}).  In particular, the relation between the two Nakanishi
 weight functions is given by
 \be
g^+(\gamma',z',z;\kappa^2,z_i)= \nonu =i\int_0^1 {d\alpha\over \alpha^3}~
 \int_{-1}^1 d\zeta' ~\widetilde{\cal G}^+({\gamma' \over \alpha} ,{z'\over
 \alpha},{\zeta'\over a};\kappa^2,z_i)\nonu \times
\theta (\alpha -|z'| -|\zeta'|)
\theta (1 -\alpha -| \zeta'-z -z' z_i|) \, . 
 \ee
with all the constrains on the variables  explicitly written.
 Indeed, notice that   the dependence upon $z$ in the weight function
$g^{(+)}_{(Ld)}(\gamma',z',z;\kappa^2,z_i)$ should be read as $z+z'z_i$ 
 (cf
 Eq. (66) in \cite{FSV1} and   \ref{distwf} of the present paper).
 From  Eq. \eqref{psidist}, the analogy with the non relativistic case appears evident, once  
  the familiar form of the global propagation is recognized. As a matter of fact, one has 
($1>z^2$)
\be {1 \over \gamma +z^2 m^2+(1-z^2)\kappa^2-i\epsilon}=\nonu =~ {(1-z^2) \over 4} ~{ 1\over M^2_0
-M^2-i\epsilon}  \, ,
\label{freepro}\ee
where $M_0$ is the free mass of the two-body system  given by
\be
M^2_0=~4~{(m^2 + \gamma )\over (1-z^2)} \, .
\ee
It should be pointed out that the cut  in $\psi_{dist}$  is mirrored in  
the integral equation determining the Nakanishi weight
function,  in particular in the part governed by the dynamics (see 
Eq. \eqref{scatlad1}, below). It is useful to anticipate that
the  cut is canceled by the proper
factor in the evaluation of the scattering amplitude.

 An issue of fundamental relevance related to    
$\psi_{dist}$ in Eq. \eqref{scat3db} (or 
 to  
$\psi_{dist}$ in Eq. \eqref{psidist}) is to determine 
the support  of the Nakanishi weight function $g^{(+)}(\gamma',z',z;\kappa^2,z_i)$ 
(or equivalently $\widetilde{\cal G}^+(\gamma',z',\zeta^\prime;\kappa^2,z_i)$)
 with respect to the non compact variable $\gamma'$. While the variable 
$\gamma=k^2_\perp$ in $\psi_{dist}\left(z,\gamma;\kappa^2,z_i\right)$ is such
that  $\gamma\in[0,\infty)$  and the same holds 
for
$\gamma'$ in the Nakanishi weight function when the bound state is discussed 
(see \cite{FSV2}), in the case of  a scattering state 
 one has  a different interval, namely $\gamma'\in (-\infty,\infty)$. Then,
 a question rises about the width of the support when $\kappa^2 \to 0^-$,
  i.e.  the zero-energy limit 
 which we are interested in. One should expect that the relevant support of $\gamma'$ had to shrink in order 
 to match the one pertaining to a bound
 state. 
 
This can be accomplished if 
 $$ lim_{\kappa^2\to 0^-}~g^{(+)}(\gamma',z',z;\kappa^2,z_i)=0$$  for 
 $\gamma'<0$. 
 Notably, this is what happens, as shown in detail in  the following subsection.
 It should be pointed out that  such a result is relevant for what follows, 
 since we are going to consider the limit of a scattering state for 
 $\kappa^2\to 0^-$, and one could be puzzled by the abrupt
 transition   of the lower extremum for $\gamma'$ from an unbound value, for
 $\kappa^2 <-\epsilon$,   to a bound one, for the zero-energy limit.

\subsection{The support of the Nakanishi weight function for the
inhomogeneous BSE}
\label{supp}
 In order to address the support issue  above introduced, let us 
 consider the first meaningful approximation 
to Eq. \eqref{inbse}, namely the approximation where the
kernel $i~{\cal K}$ is substituted by its ladder contribution, given by 
\be
i{\cal K}^{(Ld)}(k,k_i,p) = ~i {(-ig)^2 \over (k-k')^2-\mu^2+i\epsilon} \, .
\label{ladk}\ee
First, one 
inserts the Nakanishi Ansatz for the BS amplitude, 
Eq. \eqref{ptirsc}, in  the ladder BSE. Then,   { one  can perform 
the   integration over 
$k^-$ without any approximation, and obtain}    
   the ladder inhomogeneous BSE projected onto the null-plane, i.e. 
   an integral equation  that  relates 
   $\psi_{dist}$ given by Eq. \eqref{scat3db}, to the dynamics dictated by
  the ladder kernel (see details in Ref. \cite{FSV1}). Namely, one gets
 \be
\int_{-\infty}^{\infty}d\gamma'~\int_{-1}^1 dz'~
\frac{g^{(+)}_{(Ld)}(\gamma',z',z;\kappa^2,z_i)}
{[{\cal D}_2-i\epsilon]^2}
=\nonu=
~{g^2 \over \left[\gamma +z^2m^2+(1-z^2) \kappa^2  -i\epsilon \right]}
\left[{\cal W}^{(Ld)}(\gamma,z;\kappa^2,z_i) \right.
\nonu \left.+
{1 \over 2(4\pi)^2}
 \int_{-\infty}^{\infty}d\gamma'\int_{-1}^{1}d\zeta
\int_{-1}^{1}d\zeta'
~g^{(+)}_{(Ld)}(\gamma',\zeta,\zeta';\kappa^2,z_i) \right. \nonu \left.
\int_0^\infty  {d y} ~F(y,\gamma,z;\gamma',\zeta,\zeta')\right] \, ,
\label{scatlad1}\ee  
 where ${\cal W}^{(Ld)}$ 
is 
\be
{\cal W}^{(Ld)}(\gamma,z;\kappa^2,z_i)=~
{1\over  (z-z_i)}~\nonu \times~
\left\{
{\theta(z-z_i) \over  {M^2_0\over 4} (1+z)
  -{M^2\over 4} (1+z_i) + {\mu^2+\gamma\over (z-z_i)}  -i\epsilon}
+\right.
\nonu \left.-
 { \theta(z_i-z)\over   {M^2_0\over 4}(1-z)-{M^2\over 4} (1-z_i) +
  {\mu^2+\gamma\over (z_i-z)}  -i\epsilon}
\right\} \, ,
\label{calgl}\ee
and  $F$ is 
\be
 F(y,\gamma,z;\gamma',\zeta,\zeta')=
{(1+z)^2\over (1+\zeta'-z_i\zeta)^2}\nonu \times ~
{\theta (\zeta'-z-z_i\zeta) \over \left[
{\cal D}_4(y,\gamma,z;\gamma',\zeta,\zeta';z_i)
 -i\epsilon\right]^2}+ 
{(1-z)^2\over (1-\zeta'+z_i\zeta)^2} 
\nonu \times ~
{ \theta (z+z_i\zeta-\zeta') \over
\left[{\cal D}_4(y,\gamma,-z;\gamma',\zeta,-\zeta';-z_i)
 -i\epsilon\right]^2} \, ,\nonu
\label{calf}\ee
where
\be
{\cal D}_4(y,\gamma,z;\gamma',\zeta,\zeta';z_i)=\gamma +z^2 m^2+\kappa^2(1-z^2)
\nonu +
\Gamma(y,z,z_i,\zeta,\zeta',\gamma')
  +Z(z,\zeta,\zeta';z_i)
 {M^2\over 2} z z_i
\ee
with  
\be
\Gamma(y,z,z_i,\zeta,\zeta',\gamma')= { (1+z)\over (1+\zeta'-z_i\zeta)}
\nonu \times
\left\{y{\cal A}(\zeta,\zeta',\gamma',\kappa^2)+{\mu^2\over y}
+\mu^2+\gamma'\right\} \, ,
\label{gamz}\ee
\be
Z(z,\zeta,\zeta';z_i)={ (1+z)\over (1+\zeta'-z_i\zeta)}~~\zeta \, ,\nonu
\label{Zeta}\ee
and
\be
{\cal A}(\zeta,\zeta',\gamma',\kappa^2)={\zeta'}^{2}
\frac{M^2}{4} + \kappa^2 (1+{\zeta}^2)+\gamma' \, .
\label{afac} \ee
Because of    the presence of the theta
functions in Eq. \eqref{calf}, one has 
\be
 1 \geq |\zeta| \geq |Z( \pm z,\zeta, \pm \zeta',z_i)| \, .
 \label{limitz}
\ee
It should be pointed out that Eq. \eqref{scatlad1} is relevant for the calculation of the phase shifts and,
 in the zero-energy limit, of the scattering lengths (cf Sec
\ref{scatle}).

After combining the global propagation with the denominator in ${\cal W}^{(Ld)}(\gamma,z;\kappa^2,z_i)$  and  repeating the same step for $F$ 
 (see Ref. \cite{FSV1}) one can apply  the
Nakanishi theorem  on the
uniqueness of the weight-function for an $n$-leg transition amplitude
 \cite{nak71}.   
   It should be recalled that the uniqueness theorem has been proven within a perturbative
framework, while in the present context, a non perturbative one,
 the uniqueness is conjectured and numerically checked. 
Eventually, 
one gets a new integral equation for the Nakanishi weight function, 
{{that allows us to  discuss the support issue}}, viz \cite{FSV1}
\be 
g^{(+)}_{(Ld)}(\gamma,z',z;\kappa^2,z_i)
= g^2 ~\theta(-z')
~ \delta(\gamma -\gamma_a(z'))
~\times \nonu
\Bigl\{ \theta(z-z_i) ~\theta \left[1-z +z'(1-z_i)\right]
 \nonu + \theta(z_i-z)~\theta \left[1+z +z'(1+z_i)\right]\Bigr\}+\nonu -
{g^2\over 2(4\pi)^2}~
\int_{-\infty}^{\infty}d\gamma'\int_{-1}^{1}d\zeta\int_{-1}^{1}d\zeta'
~g^{(+)}_{(Ld)}(\gamma',\zeta,\zeta';\kappa^2,z_i)\nonu
\times  ~\left [{(1+z)~\theta (\zeta'-z-z_i\zeta)\over (1+\zeta'-z_i\zeta)}
h'(\gamma,z',z,z_i;\gamma',\zeta,\zeta',\mu^2)
\right. \nonu \left . +{(1-z)\theta
(z-\zeta'+z_i\zeta)\over (1-\zeta'+z_i\zeta)}
h'(\gamma,z',-z,-z_i;\gamma',\zeta,-\zeta',\mu^2)\right]
 \nonu
\label{uniq} \ee 
where 
\be
\gamma_a(z')=z'(2\kappa^2-\mu^2)~\geq ~0 \, ,
\label{cd7}
\ee
and 
$ h'(\gamma'',z',z,z_i;\gamma',\zeta,\zeta' ,\mu^2)$  is given by
\be
h'(\gamma'',z',z,z_i;\gamma',\zeta,\zeta' ,\mu^2)=~{(1+z)\over (1+\zeta'-z_i\zeta)}
\nonu \times
\Bigl\{{\partial \over \partial \lambda}\int ^\infty_0
{dy}~\int_0^1d\xi~\delta\left[z'-  \xi Z(z,\zeta,\zeta';z_i)\right]
\nonu \times~ \delta\Bigl[
{\cal F}(\lambda,y,\xi;\gamma'',z,\zeta,\zeta',\gamma';z_i,\kappa^2,\mu^2)\Bigr]
\Bigr \}_{\lambda=0} ~~.
\label{h'}\ee
with
\be
{\cal F}(\lambda,y,\xi;\gamma'',z,\zeta,\zeta',\gamma';z_i,\kappa^2,\mu^2)=
\nonu=\gamma''-
\xi ~{ (1+z)\over (1+\zeta'-z_i\zeta)}
\nonu \times ~\left( {y^2 {\cal A}(\zeta,\zeta',\gamma',\kappa^2) +y(\mu^2+\gamma')
 +\mu^2 \over y}\right)
 - \xi\lambda 
\label{delarg}\ee
Notice that  the inhomogeneous term vanishes both at 
$z=\pm 1$ and at  $z'=-1$, as expected (cf Eq. \eqref{edge}).    Indeed, for  $z=1$,  one has
\be
\theta(-z')
~
\Bigl\{ \theta(1-z_i) ~\theta \left[ z'(1-z_i)\right]
 \nonu+ \theta(z_i-1)~\theta \left[2 +z'(1+z_i)\right]\Bigr\} \, ,
 \ee
that  vanishes.  For $z=-1$, one gets 
 \be
 \theta(-z')~\Bigl\{ \theta(-1-z_i) ~\theta \left[z'(1-z_i)\right]
 \nonu + \theta(z_i+1)~\theta \left[z'(1+z_i)\right]\Bigr\} \, ,
 \ee
and again the theta functions produce a vanishing outcome. Finally, if $z'=-1$, the inhomogeneous term is  
vanishing, since \be
\theta(z-z_i) ~\theta \left[1-z -(1-z_i)\right]
\nonu + \theta(z_i-z)~\theta \left[1+z -(1+z_i)\right]=0 \,.
\ee

The integral equation based on the uniqueness theorem (that has been numerically
verified for the bound states case in Ref. \cite{FSV2}, and for the zero-energy
limit in the present work, cf Sec. \ref{results}) { {leads to understand in detail 
the sharp transition of the support in $\gamma$.}}

For $\kappa^2<0$ the support is $(-\infty,\infty)$, and one can split the integral
equation  \eqref{uniq} in two coupled integral equations: one is
inhomogeneous, while the other is  homogeneous. To show this, let us introduce the
following decomposition of the weight function 
$g^{(+)}_{(Ld)}(\gamma,z',z;\kappa^2,z_i)$
\be
g^{(+)}_{(Ld)}(\gamma,z',z;\kappa^2,z_i)= 
\theta(\gamma)~g^{(+)}_{p;(Ld)}(\gamma,z',z;\kappa^2,z_i)\nonu+~\theta(-\gamma)~
g^{(+)}_{n;(Ld)}(\gamma,z',z;\kappa^2,z_i) \, .
\ee
Inserting such a decomposition in Eq. \eqref{uniq} one gets
\be
g^{(+)}_{p;(Ld)}(\gamma,z',z;\kappa^2,z_i)=g^2 ~\theta(-z')
~ \delta(\gamma -\gamma_a(z'))
~\times \nonu
\Bigl\{ \theta(z-z_i) ~\theta \left[1-z +z'(1-z_i)\right]
 \nonu + \theta(z_i-z)~\theta \left[1+z +z'(1+z_i)\right]\Bigr\}+\nonu -
  {g^2 \over 2(4\pi)^2}~
\theta(\gamma)~\Bigl[
\int_{0}^{\infty}d\gamma'\int_{-1}^{1}d\zeta\int_{-1}^{1}d\zeta'\nonu 
\times~H'(\gamma,z',z,z_i;\gamma',\zeta,\zeta',\mu^2)
~g^{(+)}_{p;(Ld)}(\gamma',\zeta,\zeta';\kappa^2,z_i) \nonu 
+
\int_{-\infty}^{0}d\gamma'\int_{-1}^{1}d\zeta\int_{-1}^{1}d\zeta'\nonu
\times ~H'(\gamma,z',z,z_i;\gamma',\zeta,\zeta',\mu^2)
~g^{(+)}_{n;(Ld)}(\gamma',\zeta,\zeta';\kappa^2,z_i)\Bigr] \, , \nonu
\label{uniq1} \ee
and \be
g^{(+)}_{n;(Ld)}(\gamma,z',z;\kappa^2,z_i)=
 -{g^2 \over 2(4\pi)^2}~
\theta(-\gamma)~\nonu \times ~\Bigl[
\int_{0}^{\infty}d\gamma'\int_{-1}^{1}d\zeta\int_{-1}^{1}d\zeta'\nonu
\times~H'(\gamma,z',z,z_i;\gamma',\zeta,\zeta',\mu^2)
~g^{(+)}_{p;(Ld)}(\gamma',\zeta,\zeta';\kappa^2,z_i) \nonu
+
\int_{-\infty}^{0}d\gamma'\int_{-1}^{1}d\zeta\int_{-1}^{1}d\zeta'\nonu
\times~H'(\gamma,z',z,z_i;\gamma',\zeta,\zeta',\mu^2)
~g^{(+)}_{n;(Ld)}(\gamma',\zeta,\zeta';\kappa^2,z_i)\Bigr] \, ,\nonu
\label{uniq2} \ee 
with
\be
H'(\gamma,z',z,z_i;\gamma',\zeta,\zeta',\mu^2)=\nonu=
 ~\Bigl[{(1+z)\theta (\zeta'-z-z_i\zeta)\over (1+\zeta'-z_i\zeta)}
h'(\gamma,z',z,z_i;\gamma',\zeta,\zeta',\mu^2)
\nonu +{(1-z)\theta
(z-\zeta'+z_i\zeta)\over (1-\zeta'+z_i\zeta)}\nonu
\times
h'(\gamma,z',-z,-z_i;\gamma',\zeta,-\zeta',\mu^2)\Bigr] \, .
\ee
If $\kappa^2 \to 0^-$, the off-shell kernel in the homogeneous integral equation, namely the one
 with $\gamma<0$ and $\gamma'>0$, becomes vanishing and this leads to  a system
 of
 uncoupled equations.  As a matter of fact, one has for $\kappa^2 \to 0^-$  
\be
\theta(-\gamma)
~\theta(\gamma')~H'(\gamma,z',z,z_i=0;\gamma',\zeta,\zeta',\mu^2)=\nonu=
\theta(-\gamma)
~\theta(\gamma')~{(1+z)\over (1+\zeta')}
~
\Bigl\{{\partial \over \partial \lambda}\int ^\infty_0
{dy}~\nonu
\times\int_0^1d\xi~\delta\left[z'-  \xi \zeta {(1+z)\over (1+\zeta')}\right]~
\nonu \times ~ \delta\left [
{\cal F}(\lambda,y,\xi;\gamma'',z,\zeta,\zeta',\gamma';z_i,\kappa^2=0,\mu^2) \right]
\Bigr \}_{\lambda=0}=\nonu =~0 \, ,
\label{mixterm}\ee
since the delta function is always vanishing, given $
 \gamma<0 $ and
  \be
  \xi ~{ (1+z)\over (1+\zeta')}\left( {y^2 {\cal A}(\zeta,\zeta',\gamma',\kappa^2=0) 
+y(\mu^2+\gamma') +\mu^2 \over y}\right)
\nonu+\xi ~\lambda =\nonu=
\xi ~{ (1+z)\over (1+\zeta')}\left[ {y^2 \left({\zeta'}^{2} m^2 +\gamma'\right) +y(\mu^2+\gamma') +\mu^2 \over y}
\right]\nonu+\xi ~\lambda >0 \, . \nonumber
\ee
Then, for  $\kappa^2\to 0^-$, Eq. \eqref{uniq2} becomes 
\be
g^{(+)}_{n;(Ld)}(\gamma,z',z;\kappa^2=z_i=0)=
 -{g^2 \over 2(4\pi)^2}~
\theta(-\gamma)~ \nonu
\int_{-\infty}^{0}d\gamma'\int_{-1}^{1}d\zeta\int_{-1}^{1}d\zeta'~
H'(\gamma,z',z,z_i;\gamma',\zeta,\zeta',\mu^2)
\nonu \times~g^{(+)}_{n;(Ld)}(\gamma',\zeta,\zeta';\kappa^2=z_i=0) \, .
\label{uniq3} 
\ee
The above homogeneous integral equation, valid in the zero-energy limit, is expected to have as a solution 
only $g^{(+)}_{n;(Ld)}(\gamma,z',z;\kappa^2=z_i=0)=0$,  given the freedom in choosing $g^2$ 
for  scattering states. Let us recall that for the
bound state case, where $\kappa^2\geq 0$ and $\gamma>0$, one gets a homogeneous integral
 equation and deals with an eigenvalue problem. In particular, one finds   
 a discrete spectrum for $g^2 $, once a value is assigned to
 $\kappa^2$  and $\mu$ (see, e.g., Ref. \cite{FSV2}, where the bound state case is discussed,
 within the present approach). 
 
 It is also instructive to trace the behavior of the previous coupling term when
 $\kappa^2$ approaches $0^-$. If $\kappa^2$ is different from zero, than the delta function in Eq. \eqref{mixterm} can give a finite
 contribution, since its argument can vanish. To achieve such a possibility, one must have (remind that $\gamma'>0$)
 \be
 {\cal A}(\zeta,\zeta',\gamma',\kappa^2)={M^2\over 4} {\zeta'}^2+\kappa^2(1+\zeta^2) +\gamma'~<~0 \, ,
 \ee
 since the other terms, $\mu^2+\gamma'$ and $\lambda $, always yield  a positive contribution ($\lambda$ approaches zero from
 positive values). The above constraint leads to  a volume of the  integration 
  in the space $\{\gamma',\zeta',\zeta\}$ (it is a hyperboloid), that
 shrinks to zero for $\kappa^2 \to 0^-$, viz
 $$ {M^2\over 4} {\zeta'}^2+\kappa^2\zeta^2 +\gamma'~<~-\kappa^2$$

 In conclusion,
 for scattering states in the limit
$\kappa^2\to 0^-$, the corresponding Nakanishi weight function reduces to the
component $g^{(+)}_{p;(Ld)}(\gamma,z',z;\kappa^2=z_i=0)$ and
fulfills the following inhomogeneous integral
equation
\be
g^{(+)}_{p;(Ld)}(\gamma,z',z;\kappa^2=z_i=0)=g^2 ~\theta(-z')
~ \delta(\gamma -\gamma_a(z'))
 \nonu \times~
\Bigl\{ \theta(z) \theta \left[1-z +z'\right]
 + \theta(-z)\theta \left[1+z +z'\right]\Bigr\} \nonu
 -
{g^2 \over 2(4\pi)^2}\theta(\gamma)~
\int_{0}^{\infty}d\gamma'\int_{-1}^{1}d\zeta
\nonu \times\int_{-1}^{1}d\zeta'~
H'(\gamma,z',z,z_i=0;\gamma',\zeta,\zeta',
\mu^2)\nonu \times
~g^{(+)}_{p;(Ld)}(\gamma',\zeta,\zeta';\kappa^2=z_i=0)  
\label{uniq4} \ee
This is  sharply different from the general case $\kappa^2<0$ given by Eqs. \eqref{uniq1} 
and \eqref{uniq2}.

\section{The Scattering length}
\label{scatle}
In the CM frame, the differential cross section for the 
elastic scattering of two scalars   can be written as follows
\cite{zuber}
\be
{d\sigma \over d\Omega}={1 \over 64 \pi^2 s}~ |T^{el}_{ii}|^2=|f^{el}(s,\theta)|^2 \, ,
\ee
with $T^{el}_{ii}$ the invariant matrix element of the T-matrix, that is
dimensionless (recall that   in a $\phi^3$
theory the coupling constant $g$ has the dimension of a mass), and 
$f^{el}(s,\theta)$ the elastic scattering amplitude. It turns out that 
\be
f^{el}(s,\theta)= ~-{1 \over 8 \pi \sqrt{s}}~ T^{el}_{ii} \, ,
\ee
where $s=M^2$ and
  $cos\theta=\hat k_f \cdot \hat k_i$.
To introduce the relation with the phase shifts $\delta_\ell$, let us expand the
scattering amplitude on the basis of the Legendre polynomials, $P_\ell(cos\theta)$,
 as follows
\be
f^{el}(s,\theta)= {1\over k_r}~\sum_{\ell}
(2\ell+1)~ f^{el}_\ell~P_\ell(cos\theta) \, ,
\label{f00}\ee
where the relative three-momentum is $k_r=\sqrt{s/4-m^2}$, or 
$k^2_r=-\kappa^2$,
 and the projected amplitudes
are given by
\be
f^{el}_\ell=e^{i \delta_\ell} ~sin\delta_\ell \, .
\ee
Finally, in the zero-energy limit, only the amplitude with $\ell=0$ survives 
and one obtains 
\be
f^{el}_{0}\simeq \delta_0\simeq-a~k_r
 \, ,\ee
where $a$ is the s-wave {\em scattering length}. Therefore, in the zero-energy limit
one gets
\be \lim_{s \to 4m^2}f^{el}(s,\theta)= -a \ee
On the other hand, the scattering amplitude can be   calculated through the BS
amplitude as follows  (see Ref. \cite{FSV1} for details) 
\be
f^{el}(s,\theta)= -{i \over \sqrt{s}~8 \pi}~
\lim_{k'\to k_f}~\langle k',p|G_0^{-1}(p)|\Phi^{(+)};p,k_i\rangle=
\nonu=
{1 \over \sqrt{s}~8~\pi}~\lim_{(\gamma,z) \to (\gamma_f,z_f)}
\left[\gamma +z^2m^2+(1-z^2)\kappa^2  -i\epsilon
\right]\nonu \times~{4 \over (1-z^2)} ~\psi_{dist}(z,\gamma;\kappa^2,z_i) \, ,
\label{f0}\ee
where $k'=(p'_1-p'_2)/2$,  $p'_1+p'_2=p$ (recall that
$p^2=M^2=s$). 
 
 In ladder approximation and choosing $\gamma_i=0$, from Eqs. \eqref{scat3db}
 and
 \eqref{scatlad1} one gets
 \be f^{el}_{(Ld)}(s,\theta)= ~{2\alpha ~m^2\over \sqrt{s} } ~
 \Bigl[ {\cal W}^{(Ld)}(\gamma_f,z_f;\kappa^2,z_i) ~
\nonu +{1 \over 2 (4 \pi)^{2}}~
\int_{-\infty}^{\infty}d\gamma''\int_{-1}^{1}d\zeta\int_{-1}^{1}d\zeta'
\nonu \times ~ \int_{0}^\infty  {d y}~
 F(y,\gamma_f,z_f;\gamma'',\zeta,\zeta')
~g^{(+)}_{(Ld)}(\gamma'',\zeta,\zeta';\kappa^2,z_i)\Bigr] \, ,\nonu
\label{f0lad} \ee
where   $$\alpha={g^2 \over m^2 16 \pi}~~~~~~.$$
If $z \to \pm 1$ the free mass $M^2_0 \to \infty$, and one can see that
 ${\cal W}^{(Ld)}(\gamma,z=\pm 1;\kappa^2,z_i) \to 0$, by taking into account
 also  the constraints generated by
 the theta functions. 
 In general, the  denominators in Eq. \eqref{calgl}
do not vanish, since    (i) only  minus components of {\em on-mass-shell}
particles are present there, and  (ii) the momentum conservation law does not 
hold for those components (one can also 
explicitly check that the denominators do not have real roots).
Moreover, since $\gamma_i=0$ then
$M^2=m^2/z_i=(m^2+\gamma_f)/z_f$.   It is useful to introduce  some 
kinematical relations relevant for describing the scattering process.
In particular, the
initial and final Cartesian three-momenta, $\vec k_i$  and  
$\vec k_f$, has to be completed giving the third components, viz
\be
k_{iz}= {1 \over 2} \left( k^+_i -k^-_i\right)=-z_i~ {M\over 2} ~~, \nonu 
k_{fz}= {1 \over 2} \left( k^+_f -k^-_f\right)=-z_f~ {M\over 2} \, ,
\ee
Then, one can write down the relation between the scattering angle $\theta$ 
and the LF variables $z_f$ and $z_i$, given by 
\be \vec k_i \cdot \vec k_f = z_i z_f {M^2\over 4}=-\kappa^2~cos \theta \, ,
\ee
where $p\cdot k_i= p\cdot k_f=0$ has been used (those constraints are imposed by the 
on-mass-shellness of the particles in the elastic channel). It also follows that  
$$|\vec k_i|^2=|\vec
k_f|^2=-\kappa^2~~~~~.$$
Finally, by exploiting the relation $\kappa^2 =-z^2_i M^2/4$, 
that holds for $\gamma_i=0$,
one gets
\be
z_f=z_i~cos\theta \, ,
\ee
For $\kappa^2=(m^2 -s/4) \to 0^-$, both $z_i$ and $z_f$ vanish (as well as $\gamma_f=|k_{f\perp}|^2$), and 
one loses the dependence upon the scattering angle $\theta$ in the scattering amplitude, namely one has a s-wave
scattering, as it must be. The two functions,   ${\cal W}^{(Ld)}(\gamma_f,z_f;\kappa^2,z_i)$  
and $ F(y,\gamma_f,z_f;\gamma'',\zeta,\zeta')$, 
 become
\be
\lim_{\kappa^2\to 0^-}{\cal W}^{(Ld)}(\gamma_f,z_f=z_i~cos\theta;\kappa^2,z_i) = \nonu
=\lim_{\kappa^2\to 0^-}{1\over  (z_f-z_i)}~
\Bigl\{
{\theta(z_f-z_i) \over  {M^2\over 4} (z_f-z_i)
   + {\mu^2+\gamma_f\over (z_f-z_i)}  -i\epsilon}
\nonu +
 { \theta(z_i-z_f)\over  - {M^2\over 4}(z_i-z_f) -
  {\mu^2+\gamma_f\over (z_i-z_f)}  +i\epsilon}
\Bigr\}
= ~
{1\over  \mu^2} \, ,
\ee
and  
\be
\lim_{\kappa^2\to 0^-} F(y,\gamma_f,z_f=z_i~cos\theta;\gamma'',\zeta,\zeta')=
\nonu ={y^2 \over \left[
y^2~(m^2{\zeta'}^2 +\gamma'')
+y~\left(\mu^2+\gamma''
 \right)
+\mu^2-i\epsilon\right]^2} \, ,\nonu
\ee
 Then, in the zero-energy limit  Eq. \eqref{f0lad}
reduces to
(see also  \ref{zerol})
\be
\lim_{s \to 4 m^2}f^{el}_{0Ld}(s,\theta)= ~-a~
=\nonu = ~m~\alpha~
\left\{{1 \over \mu^2 }  ~+{1 \over 2 (4\pi)^{2}}~
\int_{0}^{\infty}d\gamma''\int_{-1}^{1}d\zeta'
~g^{(+)}_{0Ld}(\gamma'',\zeta') \right. \nonu \left. \times ~\int_0^\infty  d y~
 {y^2 \over \left[
y^2{\cal A}_0(\zeta',\gamma'' )
+y~\left(\mu^2+\gamma''
\right) 
+\mu^2-i\epsilon\right]^2}
\right\} \, ,\nonu
\label{f0lad1}
\ee
where the first term in the curly brackets leads to the scattering length in Born approximation, viz 
\be
a_{BA}= -~m ~{\alpha \over \mu^2} \,
\label{aborn}\ee
Moreover, $g^{(+)}_{0Ld}(\gamma'',\zeta')$ is the Nakanishi weight function in the zero-energy limit.
It can be obtained by solving two different integral equations as discussed in 
detail in 
\ref{zerol}, where   the whole matter is presented in a substantially simpler way than the one in
 Ref. \cite{FSV1}  (notice that a mistyping in Eq. (103) of \cite{FSV1} has been fixed).
 In particular, the integral equation that links $\psi_{dist}$ to the dynamics governed by the BS
  kernel
 in ladder approximation is given by
\be\int_{0}^{\infty}d\gamma''
\frac{g^{(+)}_{0Ld}(\gamma'',z)}
{[\gamma+\gamma''+ z^{ 2}
m^2 -i\epsilon]^2}
= \nonu =
{g^2\over \mu^2} ~\int^\infty_{-\infty}~d\gamma''~
{\theta(\gamma'')
\over
\left[\gamma+\gamma'' +z^2m^2   -i\epsilon\right]^2 } ~\times\nonu
\Bigl\{ \theta(z) ~\theta \left[1-z -\gamma''/\mu^2\right]
 + \theta(-z)~\theta \left[1+z  -\gamma''/\mu^2\right]\Bigr\} \nonu -
{g^2 \over 2(4\pi)^2}~
\int_{0}^{\infty}d\gamma'\int_{-1}^{1}d\zeta'~g^{(+)}_{0Ld}(\gamma',\zeta')
~\int^\infty_{-\infty}
 d\gamma'' ~\times\nonu 
~{1\over \left[\gamma +\gamma''+z^2 m^2 
-i\epsilon   \right]^2}~\theta(\gamma'')~\times
 \nonu\Bigl [{(1+z)\over (1+\zeta')}
~\theta (\zeta'-z)~{h}_0'(\gamma'',z;\gamma',\zeta',\mu^2)
 \nonu +{(1-z)\over (1-\zeta')}
~\theta
(z-\zeta')~{h}_0'(\gamma'',-z;\gamma',-\zeta',\mu^2)\Bigr] 
\label{zerom1}\ee 
Notably, ${h}_0'(\gamma'',z;\gamma',\zeta',\mu^2)$ is
the proper  kernel for a bound state with vanishing energy, as one can check in   Ref. \cite{FSV2}.

 The expression of  ${h}_0'(\gamma'',z;\gamma',\zeta',\mu^2)$ is given by
 (see details in \ref{zerol})
 \be
h^\prime_0(\gamma'',z;\gamma',\zeta',\mu^2) =\nonu =
\theta \left[-~{\cal B}_0(z,\zeta',\gamma',\gamma'',\mu^2) 
 -2 \mu \sqrt{ {\zeta'}^{2} m^2 +
\gamma'} \right]
 \nonu \times
\Bigl[- {{\cal B}_0(z,\zeta',\gamma',\gamma'',\mu^2) \over
{\cal A}_0(\zeta',\gamma')~\Delta_0(z,\zeta',\gamma', \gamma'',
\mu^2 ) }~{1\over\gamma''}  \nonu + { (1+\zeta')\over (1+z)}
 \int_{y_-}^{y_+}{dy}~{y^2 \over \left[ {y}^2{\cal A}_0(\zeta',\gamma' )+
  y(\mu^2 +\gamma')+\mu^2
\right ]^2}\Bigr]
\nonu
-{ (1+\zeta')\over (1+z)}
\int_0^{\infty}{dy}~{y^2 \over \left[ {y}^2{\cal A}_0(\zeta',\gamma' )+ y(\mu^2 +\gamma')+\mu^2
\right ]^2} \, , \nonu
\label{gzerom}\ee
with 
\be
{\cal A}_0(\zeta',\gamma' )={\zeta'}^{2} m^2 +
\gamma'= {\zeta'}^{2} m^2 +
\gamma'~> ~0~~,
\nonu
{\cal B}_0(z,\zeta',\gamma', \gamma'',\mu^2 )=
\mu^2+\gamma' -\gamma''
{ (1+\zeta')\over (1+z)}~\leq 0~~,
\nonu
\Delta^2_0(z,\zeta',\gamma', \gamma'',\mu^2 )=
{\cal B}_0^2(z,\zeta',\gamma',\gamma'',\mu^2) \nonu
- 4 \mu^2~ {\cal A}_0(\zeta',\gamma')~\geq ~0~~,
\nonu
y_\pm=
{1 \over 2{\cal A}_0(\zeta',\gamma')} \nonu \times ~
 \left[ -{\cal B}_0(z,\zeta',\gamma',\gamma'',\mu^2)
 \pm \Delta_0(z,\zeta',\gamma', \gamma'',\mu^2 )\right] \, .
\label{gzero4m}\ee

 The
  zero-energy limit of Eq. \eqref{uniq}, i.e. the integral equation based on the 
  uniqueness of the Nakanishi weight function, reads (cf Ref. \cite{FSV1,FSV2})
is \be
g^{(+)}_{0Ld}(\gamma,z)
  =
{g^2\over \mu^2}  ~\theta(\gamma)
~\theta \left[\mu^2(1-|z|)-\gamma \right] \nonu
-{g^2 \over 2(4\pi)^2}~\theta(\gamma)
  ~\int_{-1}^1 d\zeta'~\int_{0}^{\infty}
 d\gamma'~g^{(+)}_{0Ld}(\gamma',\zeta')~\times \nonu
 \Bigl[{(1+z)\over (1+\zeta')}
~\theta (\zeta'-z)~
h^\prime_0(\gamma,z;\gamma',\zeta',\mu^2)
\nonu+{(1-z)\over (1-\zeta')}
~\theta
(z-\zeta')~h^\prime_0(\gamma,-z;\gamma',-\zeta',\mu^2)
\Bigr] \, .
\label{gzero2}\ee
It should be pointed out that the presence of a non smooth behavior, like  the discontinuity around 
$\gamma \sim\mu^2(1-|z|)$, is expected if one has to reproduce the singular
behavior of the distorted part of the scattering wave function (cf Eqs.
\eqref{scat3db} and \eqref{psidist}).

As illustrated in the next Sec. \ref{results}, we have taken profit of the general structure of the weight function suggested by 
Eq. \eqref{gzero2} for 
obtaining numerical solutions of
both Eq. \eqref{zerom1} and  Eq. \eqref{gzero2}, and eventually calculating the scattering
lengths.

It is worth noting that the scattering length given by Eq. \eqref{f0lad1}
represents a  normalization for $g^{(+)}_{0Ld}(\gamma'',\zeta')~$, when $\mu\le
2m$. As a
matter of fact, from Eq. \eqref{gzero2},  one realizes that the inhomogeneous term is different from
zero only for $$0\le \gamma\le \mu^2 (1-|z|)~~~.$$
Moreover, within the previous interval and $\mu\le 2m$, the contribution to the kernel $h'_0$
that contains 
$$\theta\left[\gamma { (1\pm \zeta')\over (1\pm z)}  - \gamma' -\mu^2
 -2 \mu \sqrt{ {\zeta'}^{2} m^2 +
\gamma'} \right]$$ disappears,  since
\be
\gamma { (1\pm \zeta')\over (1\pm z)}  - \gamma' -\mu^2
 -2 \mu \sqrt{ {\zeta'}^{2} m^2 + \gamma'}~\leq 
 \nonu \leq \mu^2~{ (1-|z|)\over (1\pm z)}(1\pm \zeta') -\mu^2 - 
 2 \mu m|\zeta'|~<\nonu
 <\mu^2~(1\pm \zeta') -\mu^2 - 2 \mu m|\zeta'|<~ \mu|\zeta'|~(\pm \mu - 2 m) \, ,
\ee
The final step in the above expression is always negative when $\mu< 2m$.

 Therefore, for
 $0\le \gamma\le \mu^2 (1-|z|)$ and $\mu\le 2m$, one has
 \be
 g^{(+)}_{0Ld}(\gamma,z)
 =
{g^2\over \mu^2}  ~
~+{g^2 \over 2(4\pi)^2}~
  ~\int_{-1}^1 d\zeta'~\int_{0}^{\infty}
 d\gamma' \nonu\int_0^{\infty}{dy}~{y^2 ~g^{(+)}_{0Ld}(\gamma',\zeta')\over \left[ {y}^2{\cal A}_0(\zeta',\gamma' )+ y(\mu^2 +\gamma')+\mu^2
\right ]^2}
=
\nonu=  -16 \pi m~a \, ,
\label{normg}
 \ee
where Eq. \eqref{f0lad1} has been exploited in the last step.
 
\section{Results}
\label{results}
In this Section, the numerical studies of both the scattering length and 
the distorted part of the 
3D wave function  are presented. First of all, let us illustrate our 
 numerical
method for solving the two integral equations in \eqref{zerom1} 
and \eqref{gzero2}. The main ingredient is the following decomposition of the Nakanishi
weight function that takes into account   the singular behavior shown in  Eq.
\eqref{gzero2}, but also   the result in Eq.  \eqref{normg},
that holds for $\mu\le 2m$ (this is always fulfilled  for realistic
cases)
\be
g^{(+)}_{0Ld}(\gamma,z)=~\beta~\theta(-t)+
\theta(t)\sum_{\ell=0}^{N_z} \sum_{j=0}^{N_g}
A_{\ell j} ~G_\ell(z) ~{\cal L}_j(t) ~,\nonu
\label{bas1}\ee
where (i) $t= \gamma-\mu^2(1-|z|)$, (ii) the functions $G_\ell(z)$ are given in terms of even Gegenbauer
 polynomials, $ C^{(5/2)}_{2\ell}(z)$  (recall that $g^{(+)}_{0Ld}(\gamma,z)$ must be even in $z$)  by 
\be
G_\ell(z)= 4~(1-z^2) ~\Gamma(5/2)~\nonu \times~\sqrt{{(2\ell+5/2) ~(2\ell)! \over \pi
 \Gamma(2\ell+5)}}~ C^{(5/2)}_{2\ell}(z) \, ,\nonu
\label{bas2}\ee
and (iii)
the  functions ${\cal L}_j(t)$ are expressed in terms of the Laguerre polynomials,  $
L_{j}(bt)$, by
\be
{\cal L}_j(t)= \sqrt{b}~ L_{j}(bt) ~
e^{-bt/2}~~~~~.\label{bas3}\ee
The following orthonormality conditions are fulfilled
\be
\int^1_{-1}dz~G_\ell(z)~G_n(z)=\delta_{\ell n} ~~~~~,
\nonu
  \int_0^\infty~ dt~{\cal L}_j(t)~{\cal L}_\ell(t)= \nonu =
b~ \int_0^\infty~ dt~ e^{-bt}~L_{j}(bt) ~L_{\ell}(bt)=
~\delta_{j\ell} \, .
\ee
In  \ref{mdeco}, some details are given for illustrating how Eq.
\eqref{gzero2} can be numerically solved by using the previous decomposition.
In order to speed up the convergence, in the actual calculations the parameter $b=15.0/m^2$ has been adopted. The finite-range integrations (as those with respect to the variable $z$ and the variable $\gamma$ when integrated up to $\mu^2(1-|z|)$) have been performed using
a Gauss-Legendre quadrature rule. The infinite-range integrations, on the other hand, have been performed adopting a Gauss-Laguerre
 quadrature method. Finally, the convergence of the expansion given 
in Eq.\eqref{bas1} is very rapid, and adopting the values $N_z=10$ and $N_g=24$ well 
converged values have been obtained. All the results presented in this Section have been obtained for such a choice.
Notice that at the end of the calculation $\beta$ resulted to be in agreement
with
the normalization $-16\pi a$ shown in Eq. \eqref{normg}.

In Tables \ref{tab1}, \ref{tab2} and \ref{tab3}, the scattering lengths, 
Eq. \eqref{f0lad1}, calculated by
using the Nakanishi weight function obtained by solving both 
the integral equation  \eqref{zerom1}, $a_{FVS}$, and  the integral equation \eqref{gzero2}, $a_{UNI}$,
 are shown for $\mu/m=0.15,~0.5,~1.0$ and   values 
 of the coupling constant $\alpha=g^2/(16\pi m^2)$, that range from a weak-interaction regime to a strong one. Moreover, for the sake
 of  comparison, 
  the results of Ref. \cite{carbonell6}, $a_{CK}$,  evaluated within a
 totally different framework, based on a direct calculations of the
 half-off-shell scattering amplitude  taking explicitly into account
 contributions from the singularities affecting the amplitude itself, are presented in the 
 second column. 
 For reference, also the Born values
 of the scattering lengths are given in the fifth column.
 From the Tables,  one can observe a very good
agreement among all the three sets of numerical results, but some comments are
in order: (i) the comparison between  $a_{FSV}$ and  $a_{UNI}$ clearly confirms that
the uniqueness of the Nakanishi weight function can be assumed with a very 
high degree
of confidence, as we have quantitatively  shown    also for the bound-state 
case
\cite{FSV2,FSV3}; (ii)  differences between  $a_{CK}$ \cite{carbonell6} and our calculations are present for $\mu=0.15~m$
when  the value of $\alpha$ approaches  
 a value which corresponds to a
bound state of zero-energy. In such a case, the scattering length diverges (let us recall that, for the bound-state case, $\alpha$ is obtained as an eigenvalue
of the homogeneous integral  equation, in ladder
approximation), or there is a change of sign. Indeed, 
the above mentioned numerical differences do not represent a big issue
{ (nonetheless it
will numerically investigated elsewhere)}, given
the completely different theoretical frameworks adopted in 
Ref. \cite{carbonell6} and in
our work, and the well-known resonance behavior  of the scattering
length,  when a bound state is approaching  a zero-energy scattering state.
Finally, it is worth noting that the Born approximation $a_{BA}$ represents a  
quite good
  approximation { only for
small $\alpha$ (see also the following Fig. \ref{a_fig}). Summarizing, the results shown in Tables \ref{tab1}, \ref{tab2} and  \ref{tab3}, together with the calculations for the bound states
\cite{carbonell1,FSV2,FSV3},  are a very strong evidence that the Nakanishi Ansatz, like the one for scattering states 
in Eq. \eqref{ptirsc}, represents a reliable tool for solving both homogeneous
and inhomogeneous
BSE's in Minkowski space.}
\begin{table}
\centering
\caption
{ Comparison, for $\mu/m=0.15$, between the scattering lengths (see Eq. \eqref{f0lad1}) 
evaluated  in Ref. \cite{carbonell6}, $a_{CK}$ (second column),  and
the  ones, $a_{FSV}$ (third column) and  $a_{UNI}$ (fourth column),  calculated by adopting  
the Nakanishi weight
function obtained from Eqs. \eqref{zerom1} and  \eqref{gzero2}, respectively. All the calculations are in ladder approximation.
The first column contains the coupling constant $\alpha=g^2/(16 \pi m^2)$.  
Finally, the fifth
column shows the scattering length in Born approximation, Eq. \eqref{aborn}.
  The scattering lengths are in unit $1/m$.($^*$Private communication by J. Carbonell)}


\begin{tabular}{r r r r r}
\hline\noalign{\smallskip}
 $\alpha$ & $a_{CK}$  \cite{carbonell6}&  $a_{FSV}$  & 
 $a_{UNI}$ & $a_{BA}$\\
 \noalign{\smallskip}
 \hline\noalign{\smallskip}
   0.01 & -0.460    & -0.459  &  -0.459   & -0.444 \\ 
  0.05 & -2.70     & -2.65   &  -2.65    & -2.22 \\   
  0.10 & -6.92     & -6.66   &  -6.66    & -4.44  \\  
  0.20 & -34.6     & -29.9   &  -29.9    & -8.89  \\  
  0.30 &  79.5     &  105.9  &   105.0   & -13.3 \\   
  0.40 &  27.2     &  28.0   &   28.0    & -17.8 \\   
  0.50 &  17.7$^*$ &  17.2   &   17.2    & -22.2 \\   
  0.60 &  12.8     &  11.8   &   11.8    & -26.7 \\   
  0.70 &  8.66     &  7.73   &   7.72    & -31.1 \\   
  0.80 &  3.73     &  3.68   &   3.68    & -35.6 \\   
  0.90 &  -4.57    & -1.13 & -1.13  & -40.0 \\        
  1.00 & -28.1     & -7.89 & -7.89  & -44.4 \\         
  1.10 &  900.     & -19.4     &   -19.4    & -48.9 \\ 
  1.50 &  24.7     &  66.9    &    66.9    & -66.7\\   
  2.00 &  17.4     &  23.2    &    23.2    & -88.9\\   
  2.50 & 14.4      &  12.3    &    12.2    & -111.0\\  \noalign{\smallskip} \hline
\end{tabular}

\label{tab1}
\end{table}
\begin{table}
\centering
\caption{ The same as in Tab. \ref{tab1}, but for the mass of the exchanged scalar $\mu/m=0.5$.}

\begin{tabular}{r r r r r}
\hline\noalign{\smallskip}
 $\alpha$ & $a_{CK}$  \cite{carbonell6}&  $a_{FSV}$ & $a_{UNI}$ & $a_{BA}$ 
 \\\noalign{\smallskip}
 \hline\noalign{\smallskip}
  0.01 & -0.0403  & -0.0403&  -0.0403 &-0.04\\ 
  0.05 & -0.209   & -0.209 &  -0.209  & -0.20\\
  0.10 & -0.438   & -0.438 &  -0.438  &-0.40\\ 
  0.20 & -0.971   & -0.971 &  -0.971  &-0.80\\ 
  0.30 & -1.64    & -1.64  &  -1.64   &-1.20\\ 
  0.40 & -2.50    & -2.50  &  -2.50   &-1.60\\ 
  0.50 & -3.66    & -3.66  &  -3.66   &-2.00\\ 
  0.60 & -5.34    & -5.34  &  -5.34   &-2.60\\ 
  0.70 & -7.98    & -7.99  &  -7.98   &-2.80\\ 
  0.80 & -12.8    & -12.8  &  -12.8   &-3.20\\ 
  0.90 & -24.7    & -24.7  &   -24.8  &-3.60\\ 
  1.00 & -103.0   & -103.2  &   -103.0 &-4.00\\ 
  1.10 &  62.0    &  61.9  &    61.8  &-4.40\\  
  1.50 &  11.0    &  11.0   & 11.0     & -6.00\\ 
  2.00 &  6.34    &  6.34   &  6.34    &-8.00\\  
  2.50 &  4.54    &  4.53   & 4.53     &-10.00\\ \noalign{\smallskip}\hline
\end{tabular}
\label{tab2}
\end{table}
\begin{table}
\centering
\caption{ The same as in Tab. \ref{tab1}, but for the mass of the exchanged scalar $\mu/m=1.0$.}

\begin{tabular}{r r r r r}
\hline\noalign{\smallskip}
 $\alpha$ & $a_{CK}$  \cite{carbonell6}&  $a_{FSV}$  & $a_{UNI}$  & 
 $a_{BA}$ \\\noalign{\smallskip}
 \hline\noalign{\smallskip}
   0.01 & -0.010 & -0.010 & -0.010 & -0.01\\
  0.05 & -0.051 & -0.051 & -0.051 & -0.05\\ 
  0.10 & -0.104 & -0.104 & -0.104 & -0.10\\ 
  0.20 & -0.217 & -0.217 & -0.217 & -0.20\\ 
  0.30 & -0.339 & -0.340 & -0.340 & -0.30\\ 
  0.40 & -0.474 & -0.474 & -0.474 & -0.40\\ 
  0.50 & -0.621 & -0.621 & -0.621 & -0.50\\ 
  0.60 & -0.784 & -0.784 & -0.784 & -0.60\\ 
  0.70 & -0.965 & -0.966 & -0.966 & -0.70\\ 
  0.80 & -1.17  & -1.17  & -1.17  & -0.80\\ 
  0.90 & -1.40  & -1.40  & -1.40  & -0.90\\ 
  1.00 & -1.66  & -1.66  & -1.66  & -1.00\\ 
  1.10 & -1.95  & -1.96  & -1.96  & -1.10 \\
  1.50 & -3.79  & -3.80  & -3.81  & -1.50\\  
  2.00 & -11.1  & -11.3  & -11.3  & -2.00 \\ 
  2.50 &  56.8  &  51.7  &  51.6  & -2.50\\ \noalign{\smallskip}\hline
\end{tabular}
\label{tab3}
\end{table}

{ In Fig. \ref{a_fig}, the scattering lengths for the above three values 
of $\mu/m$
 are presented as a function of  the absolute value of the scattering length
in Born approximation $|a_{BA}|$ (see Eq. \eqref{aborn} and the last columns
 in  Tables \ref{tab1},  \ref{tab2} and \ref{tab3}). Interestingly, in the same
 figure, it is also shown the comparison with  the corresponding 
 non relativistic scattering lengths, evaluated through a well-known  
 expression (see e.g. \cite{Wein,HK}), that exactly reproduce the  second Born approximation, viz 
 \be
m~ a = {m~a_{BA}\over 1 +{\mu\over 2m} (m a_{BA})}~~~~.
 \label{anr}
 \ee
 The chosen range of $|a_{BA}|$ is
  $[0,1.5]$, in unit of the inverse mass $m$ (the mass of the interacting
  scalars). Beyond this interval, the scattering lengths can
  change the sign, as illustrated by the above Tables. Moreover, since
  $m~ a_{BA} = -\alpha ~m^2/\mu^2$, after fixing the  value of $\mu/m$ 
  one can follows
  the dependence of the scattering length on  the Yukawa coupling constant
  $g^2$.
   In particular,  from Fig. \ref{a_fig}, one can see that for increasing values of $g^2$ and
  $\mu/m$ 
   the relativistic treatment  in Minkowski
  space, becomes more and more important, as expected, since the effect of the attractive interaction  becomes more and more 
  large. Notice that the scalar exchange in Eq. \eqref{ladk} leads to a non relativistic attractive Yukawa potential. Summarizing, 
  modulo the
  adopted ladder approximation, the comparison suggests
  that some care should be taken when one has to consider the effect of the interaction in the description of both 
  hadronic scattering processes,
  even in the low-energy regime, and  final states, that, e.g., are relevant for hadronic decays.
 }
\begin{figure}
\includegraphics[width=9.2cm] {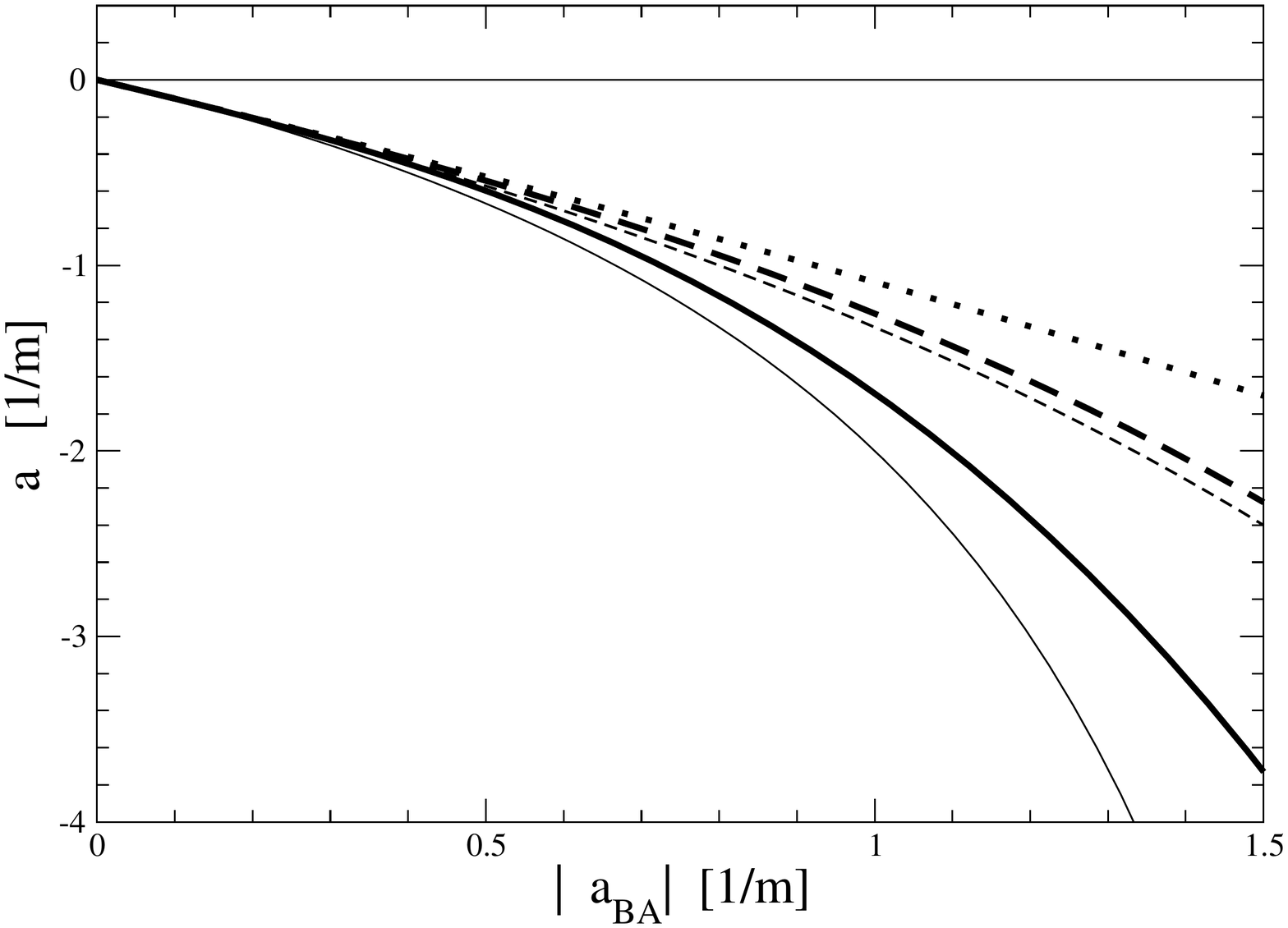}
\caption{The scattering lengths, calculated by using Eq. \eqref{f0lad1}, i.e.
corresponding to solutions of the inhomogeneous BSE at zero-energy, vs
$|a_{BA}|$ (Eq. \eqref{aborn}). Thick-solid line: $a$ for
$\mu/m=1$. Thick-dashed line: $a$ for
$\mu/m=0.5$. Thick-dotted line: $a$ for
$\mu/m=0.15$. The non relativistic scattering lengths, represented by the 
corresponding thin lines, have been calculated by using Eq. \eqref{anr}. 
Notice that for $\mu/m=0.15$ the non relativistic calculation largely 
overlaps  the relativistic one (thick-dotted line), and therefore it is 
indistinguishable.}
\label{a_fig}\end{figure}

In Fig. \ref{fig1},     the Nakanishi weight 
functions for (i) $\mu/m =0.15,~0.5,~1.0$, (ii) $\alpha=0.1,~2.5$, 
and    (iii) $z=0$,
but running $\gamma/m^2$,
are shown. It should be pointed out that, for each value of 
$\mu/m$, the 
two values of the coupling constant $\alpha$ are representatives of  a weak-interaction regime and   a 
strong one. Moreover, since the Nakanishi weight functions obtained from Eq. \eqref{zerom1} and Eq. \eqref{gzero2} 
substantially coincide, only the
calculations corresponding to Eq. \eqref{zerom1}  are shown. As mentioned at the beginning of this Section,  the step-function behavior for small $\gamma$ has to be present, and the
discontinuities are needed for obtaining the expected singularities in $\psi_{dist}$, like the one due to the global propagation. 
In Fig.\ref{fig1}, the transition from the weak regime to the strong one  increases the discontinuous behavior, that for large $\mu$ become more and
more smooth. Finally, recalling that for a bound state and $\mu\to 0$, i.e. the Wick-Cutkosky model \cite{GW,Cut}, the Nakanishi weight  
 function becomes proportional to $\delta(\gamma)$, it is instructive to see the onset of such a behavior in the upper part of  Fig.\ref{fig1}.

For $\gamma=0$ and  $|z|\ne 1 $, only the first part of the decomposition in
 Eq. \eqref{bas1}, i.e.
$g^{(+)}_{0Ld}(\gamma,z)\sim~\beta~\theta(\mu^2(1-|z|))$, is dominant, and therefore trivial.

{ In Fig. \ref{fig2},   the same quantities as in Fig. \ref{fig1}, 
but for $\gamma/m^2=0.1$ 
and running $z$, are also shown. As illustrated by the figure, the Nakanishi
weight function acquaints  a quite discontinuos behaviour, as $\mu/m$ 
increases.}

 { Indeed,}  it is more profitable to present LF distributions, obtained from  the distorted part of the 
 zero-energy 3D scattering wave 
function. In analogy with the
 bound-state case (see Refs. \cite{FSV2,FSV3}), one can construct transverse and longitudinal LF momentum distributions.  In particular,  
  one gets  the following expression for $\psi^{(Ld)}_{dist}\left(z,\gamma;\kappa^2=z_i=0\right)$
\be
\psi^{(Ld)}_{dist}\left(z,\gamma;\kappa^2=z_i=0\right)= \nonu
={(1-z^2)\over  4} ~
\int_{0}^{\infty}d\gamma'\frac{g^{(+)}_{0Ld}(\gamma',z)}
{[\gamma'+\gamma+ z^{ 2} m^2
   ]^2} \, .
\label{psidis0}\ee 
It should be noticed that 
inserting in Eq. \eqref{psidis0} only the first part
of the decomposition \eqref{bas1}, one quickly reobtains the singular behavior
due to the global propagation as shown in  Eq. \eqref{psidist}, viz
\be
\psi^{(Ld)}_{dist}\left(z,\gamma;\kappa^2=z_i=0\right) \sim ~
\beta {(1-z^2)\over  4}\nonu \times ~
\left[{1 \over  \gamma+ z^{ 2} m^2
   } - {1\over \mu^2(1-|z|)+\gamma+ z^{ 2} m^2} \right]=\beta \nonu
  \times~ {(1-z^2)\over  4}
\left[{\mu^2(1-|z|) \over  (\gamma+ z^{ 2} m^2)~
    [\mu^2(1-|z|)+\gamma+ z^{ 2} m^2]} \right] \, .\nonu
\label{psidis1}\ee 
Therefore, one has to expect singularities in the LF momentum distributions, 
{ that we would introduce}
in analogy with the ones for the bound states \cite{FSV2}.
Let us emphasize, that only for the bound states they have a probabilistic
interpretation.
One could defines (i) the distorted transverse
 LF distribution 
\be
{\cal P}_{dist}(\gamma)= {1 \over 2(2 \pi)^3}~  \int_0^1 {d\xi\over 2~\xi(1-\xi)}~  
\nonu \times \int_0^{2\pi} d\phi~
 [\psi^{(Ld)}_{dist}\left(z,\gamma;\kappa^2=z_i=0\right)]^2= 
 {1 \over (16 \pi)^2}\nonu \times \int_{-1}^1 dz~ (1-z^2) 
 \left[ \int_0^{\infty}d\gamma'~
{g^{(+)}_{0Ld}(\gamma',z)
\over [\gamma'+\gamma +z^2 m^2]^2}\right]^2 \, ,\nonu
\label{probgam}
\ee
and (ii) the longitudinal one, viz
\be
\phi_{dist}(\xi)= {1 \over (2 \pi)^3}~  {1\over 2~\xi(1-\xi)}\nonu \times ~ \int 
 d{\bf k}_\perp~
 [\psi^{(Ld)}_{dist}\left(z,\gamma;\kappa^2=z_i=0\right)]^2=
~2~
  {(1-z^2) \over (16\pi)^2}\nonu \times~ \int_0^{\infty} 
 d\gamma~\left [ \int_0^{\infty}d\gamma'~{
g^{(+)}_{0Ld}(\gamma',z)\over
[\gamma'+\gamma +z^2 m^2]^2}\right]^2  \, ,\nonu
\label{phixi} 
\ee
with the fraction of longitudinal momentum  given by 
\be
\xi ={ 1-z\over 2}=~{1 \over p^+}~\left({p^+  \over 2} +k^+\right) \, .
\ee
For the sake of
presentation, it is useful to partially removing the singularities affecting the above distributions. Therefore,  in Fig. \ref{fig3}, $\gamma^2~{\cal P}(\gamma)$ and 
$|1-2\xi|^{3/2} \phi(\xi)$ are shown. Figure \ref{fig3} illustrates the 
overall behavior
of the LF distributions by varying the coupling $\alpha$ and the mass of the
exchanged scalar $\mu$, as in Figs. \ref{fig1} and \ref{fig2}. It is worth
noting the order-of-magnitude differences, when the coupling
$\alpha$ is changed, but the typical features that one expects are still
recognizable. A part the divergent behavior, already pointed out, that can be
ascribed to the global propagation, the transverse distribution shows the
ultraviolet tail produced by the dominance of a single exchanged scalar, exactly as in
the case of the corresponding distribution for bound states (see Ref. \cite{FSV3}).
As to the longitudinal distributions, the expected peak at $\xi=1/2$ or $z=0$ is also seen.

The practical use of $\psi_{dist}$ is given by  the evaluation of reactions that need a reliable treatment of the relativistic effects, i.e. when the
coupling constant becomes larger and larger.
\begin{figure*}[th]
\includegraphics[width=8.5cm] {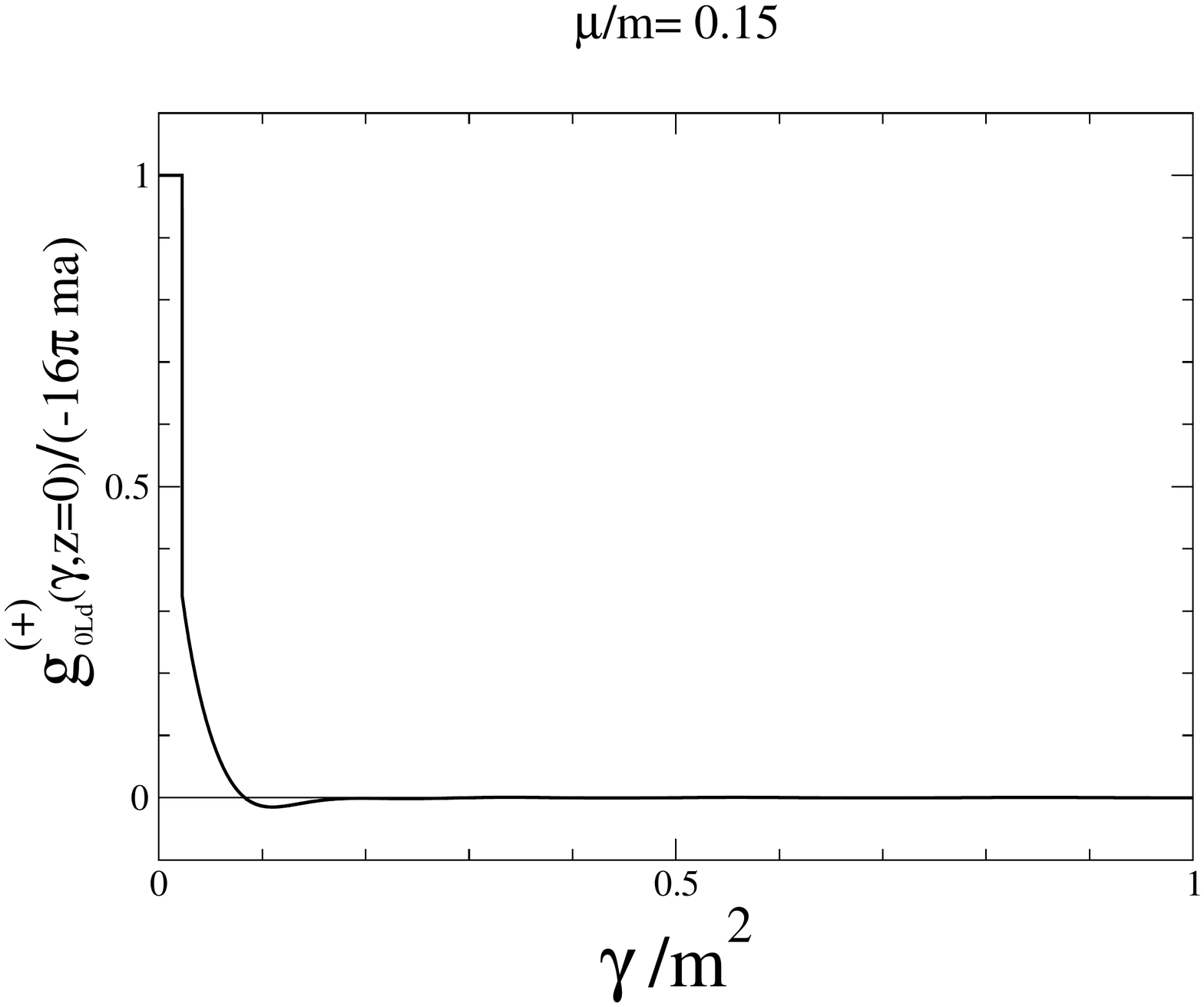}
\includegraphics[width=8.5cm] {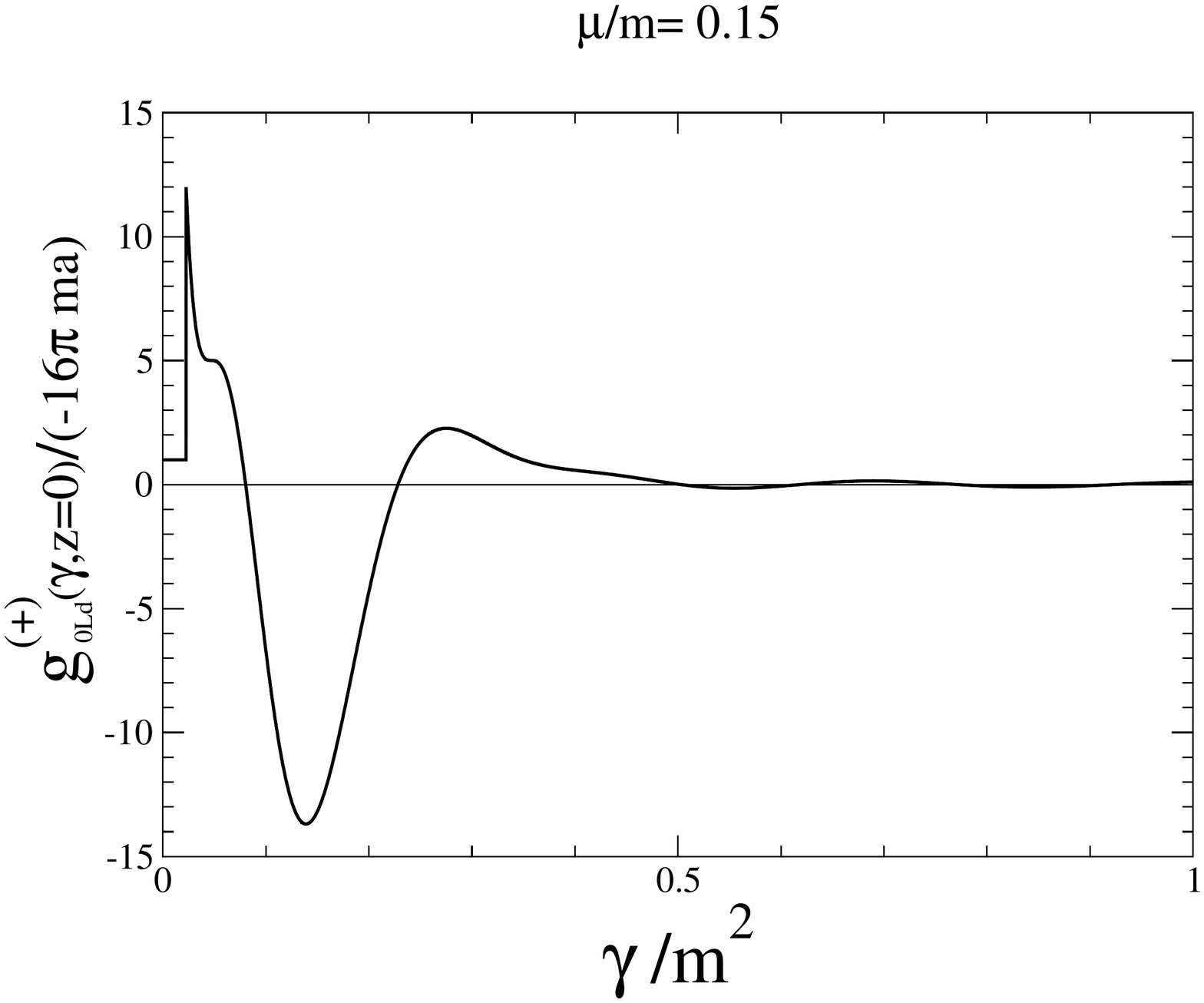}

\includegraphics[width=8.5cm] {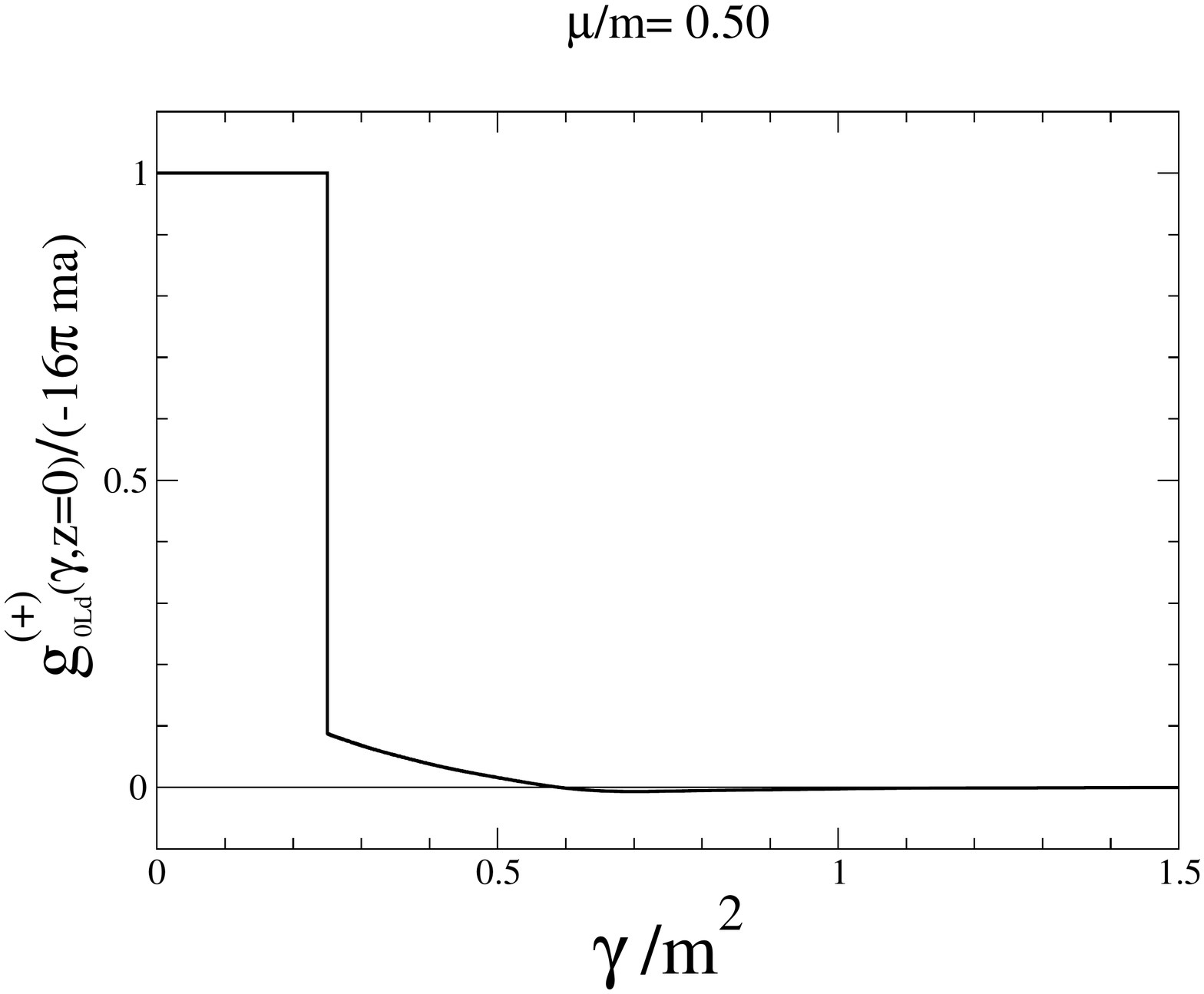}
\includegraphics[width=8.5cm] {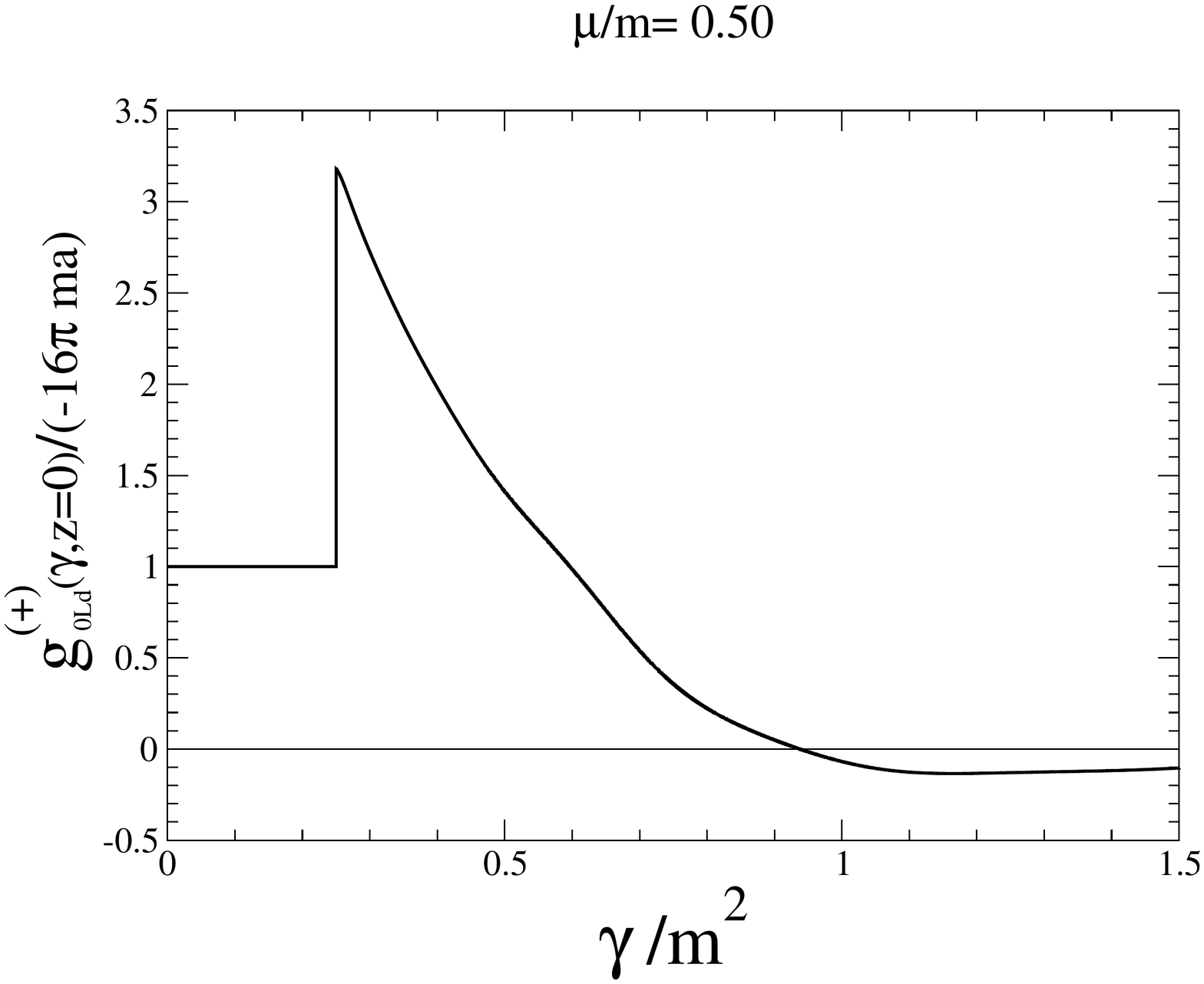}

\includegraphics[width=8.5cm] {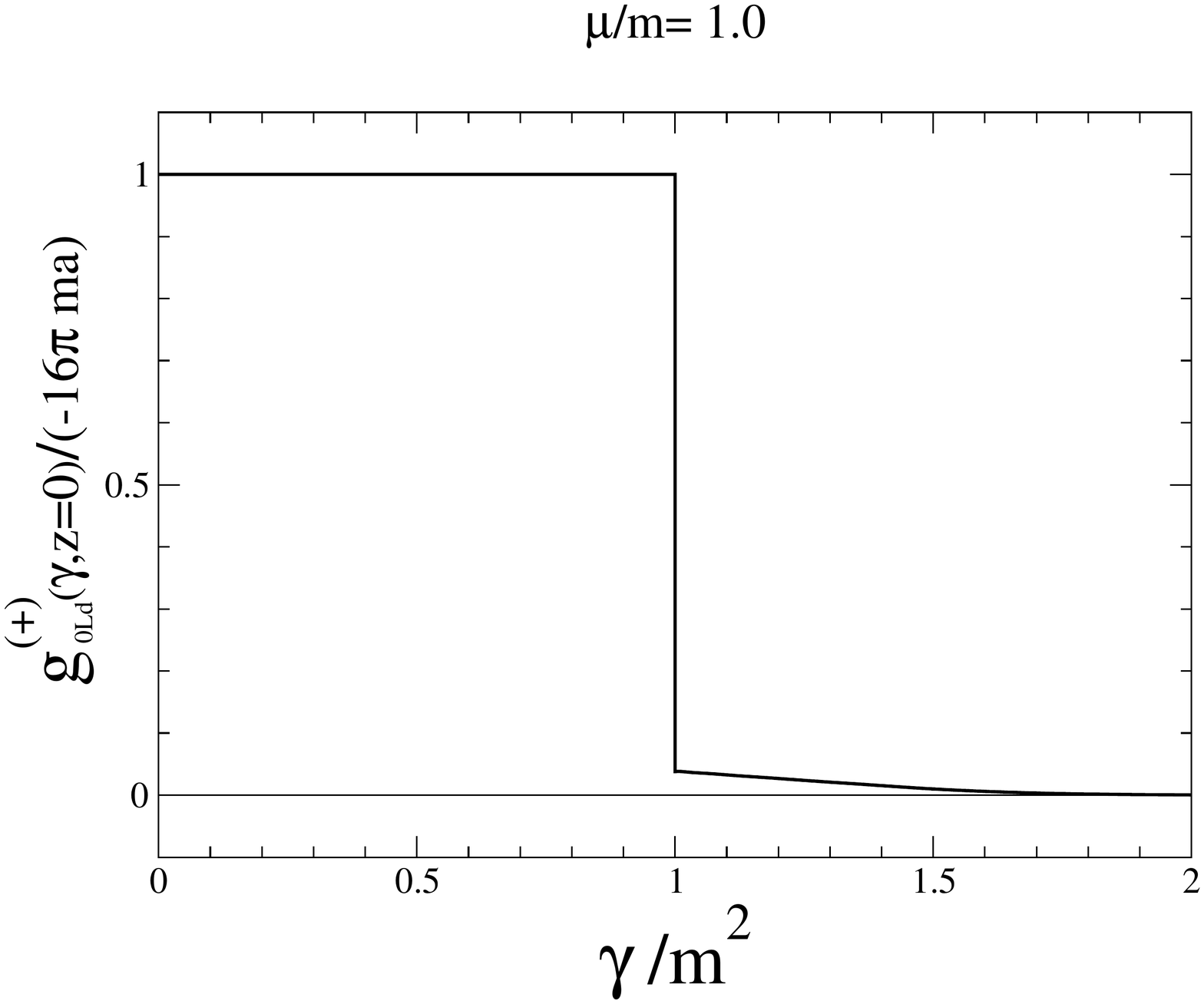}
\includegraphics[width=8.5cm] {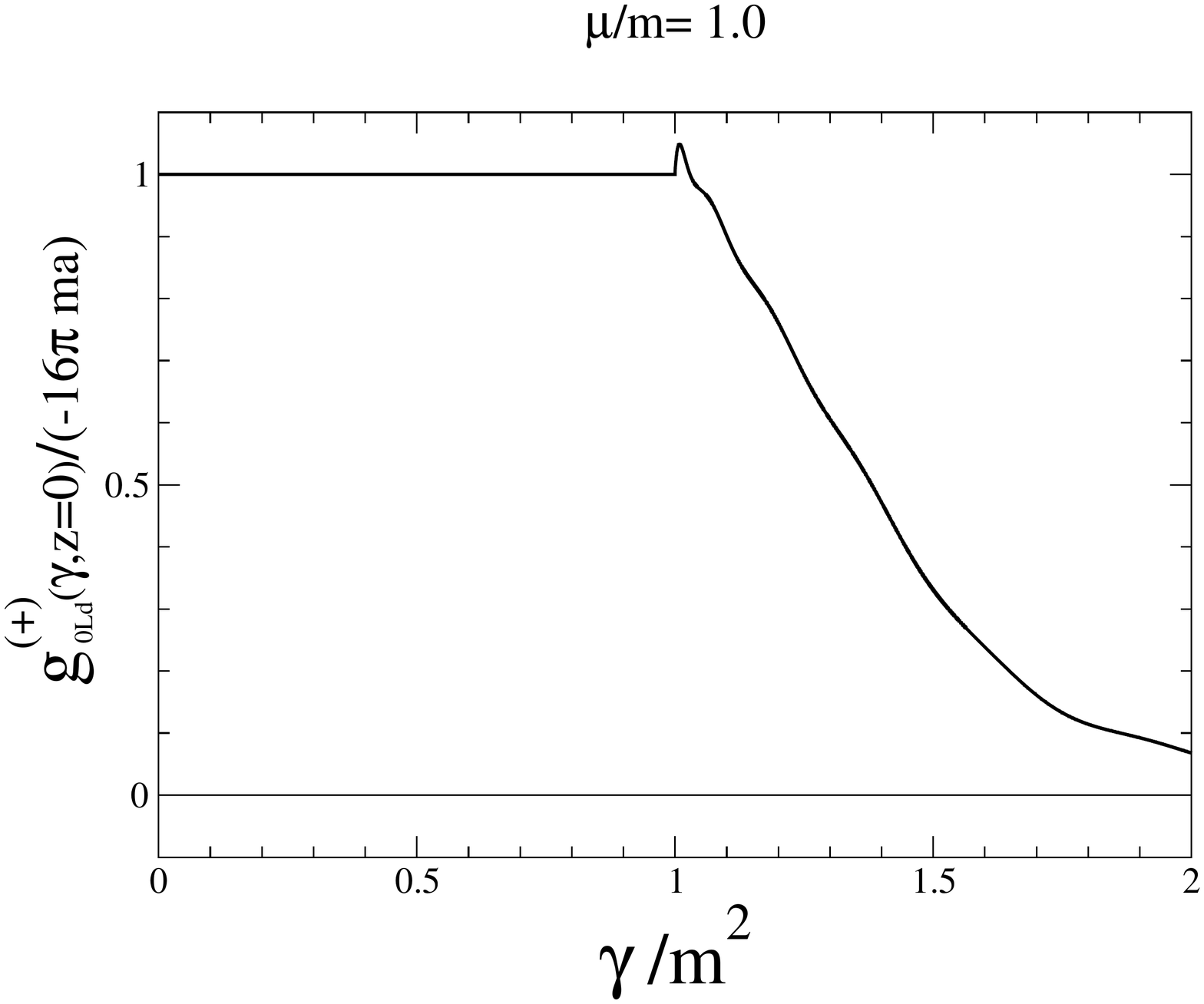}
\caption{The Nakanishi weight function $g^{(+)}_{0Ld}(\gamma,z)$, in the
zero-energy limit,   vs $\gamma/m^2$, for
$\mu/m=0.15,~0.5,~1.0$ and  $z=0$, Left panels: weak-interaction
regime with a chosen value $\alpha=0.1$. Right panels: strong-interaction
regime with a chosen value $\alpha=2.5$.}
\label{fig1}\end{figure*}

\begin{figure*}[th]
\includegraphics[width=8.5cm] {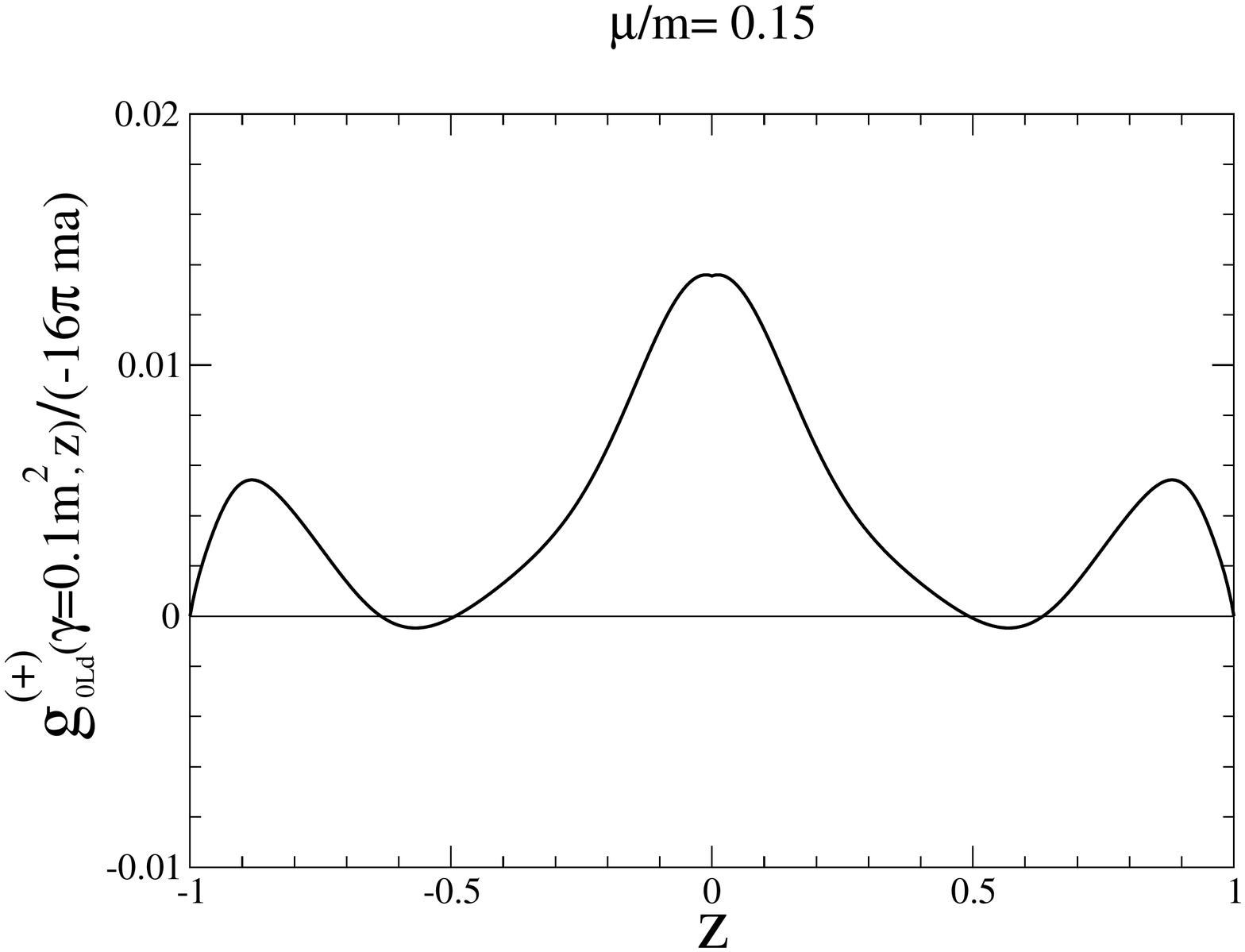}
\includegraphics[width=8.5cm] {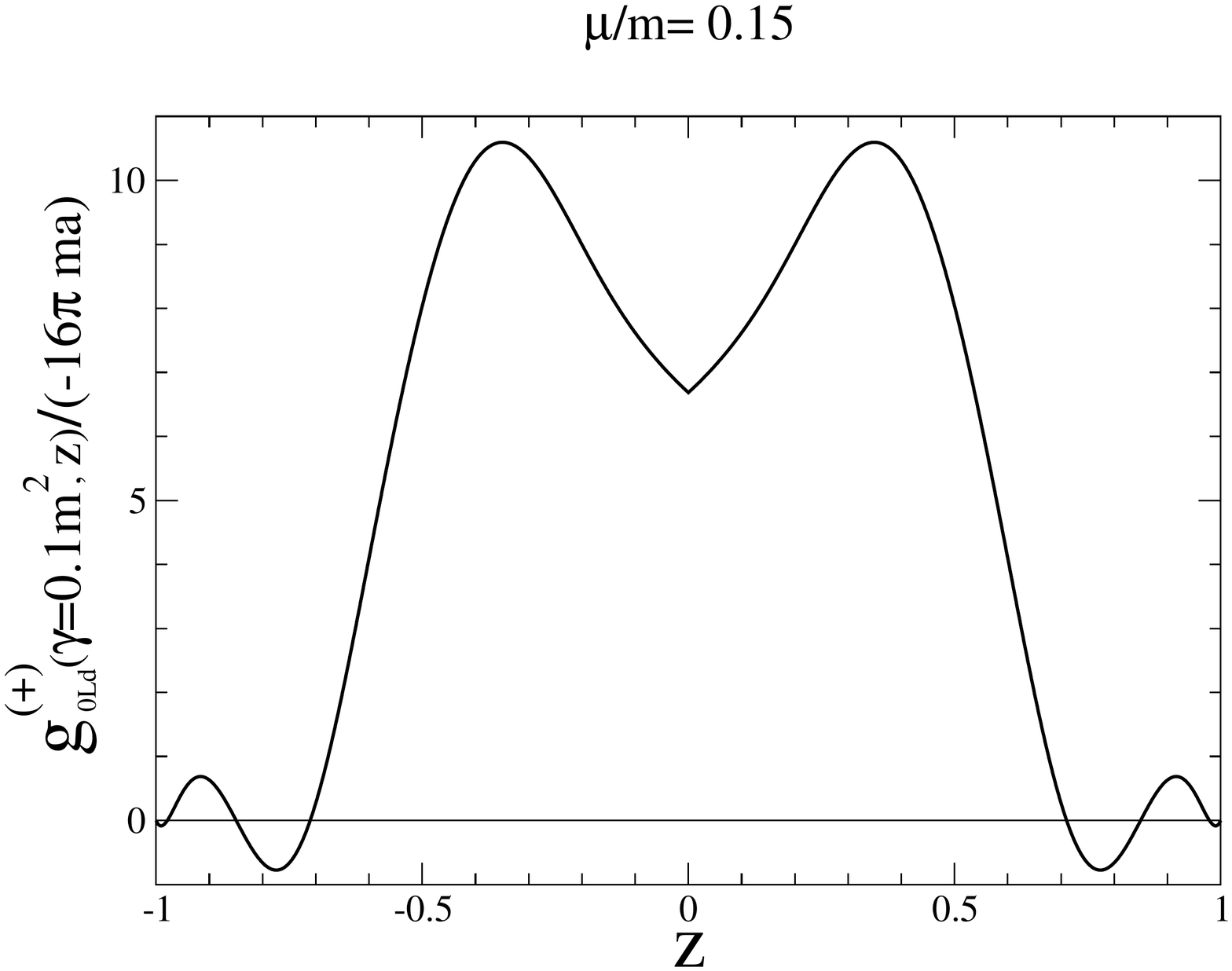}

\includegraphics[width=8.5cm] {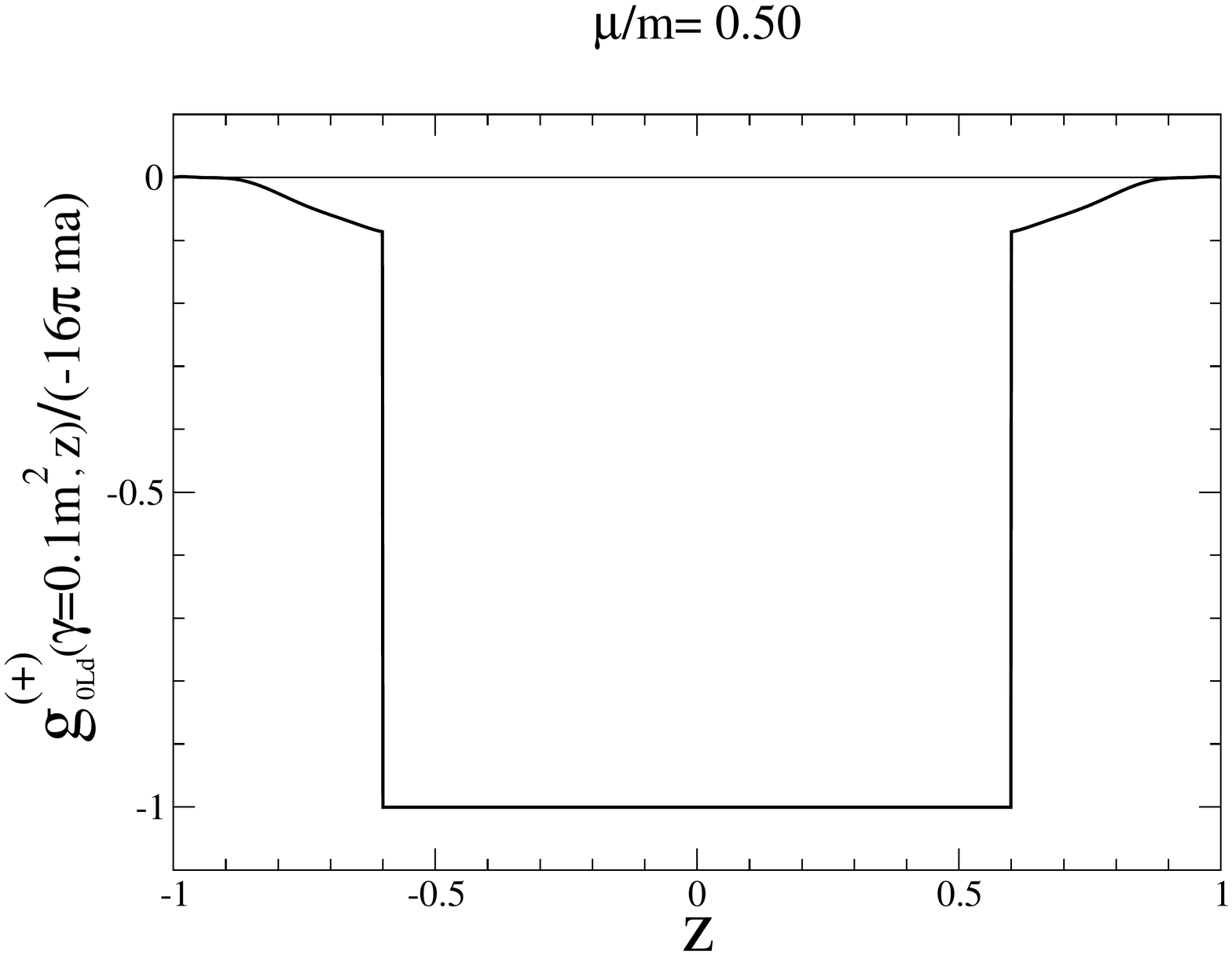}
\includegraphics[width=8.5cm] {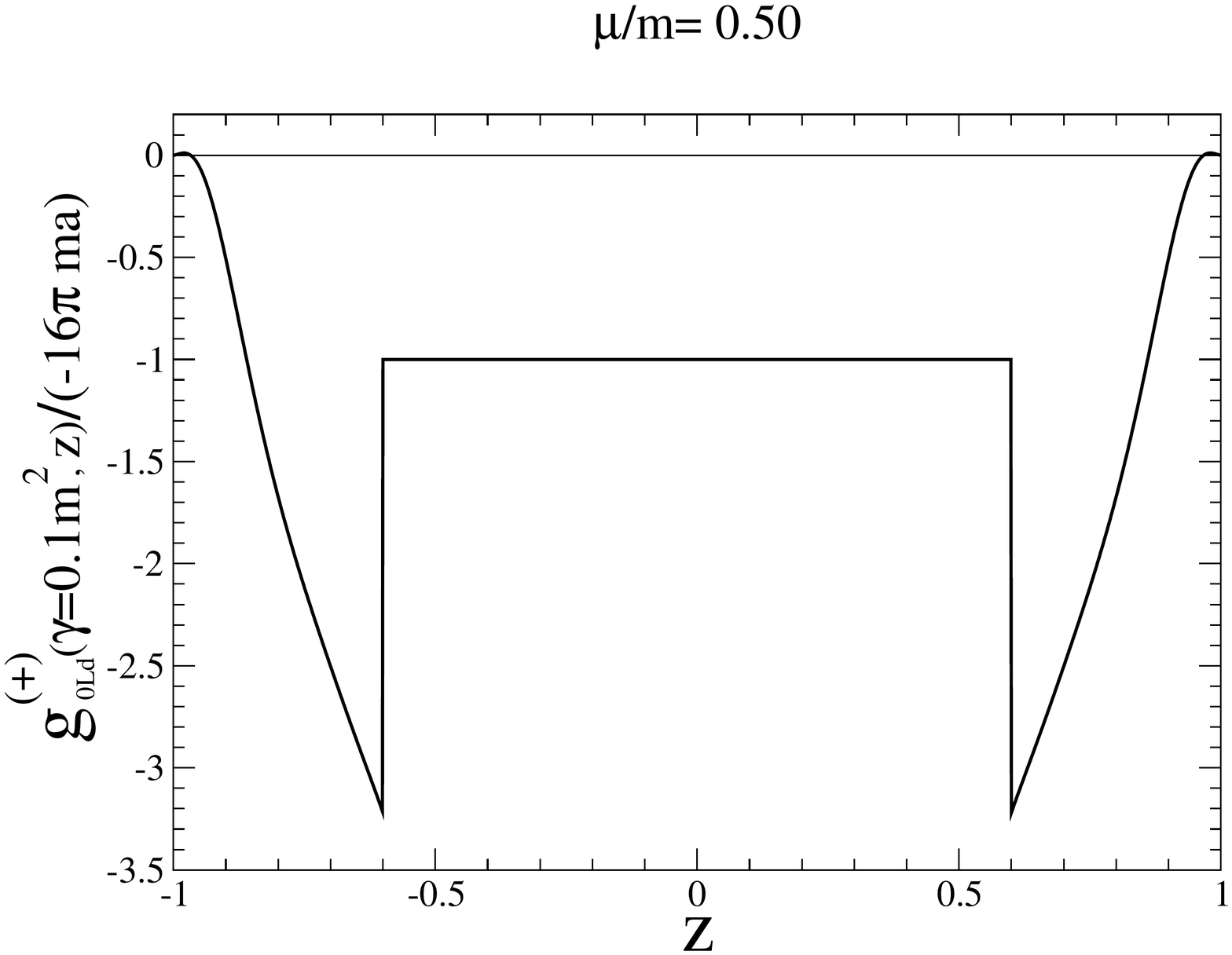}

\includegraphics[width=8.5cm] {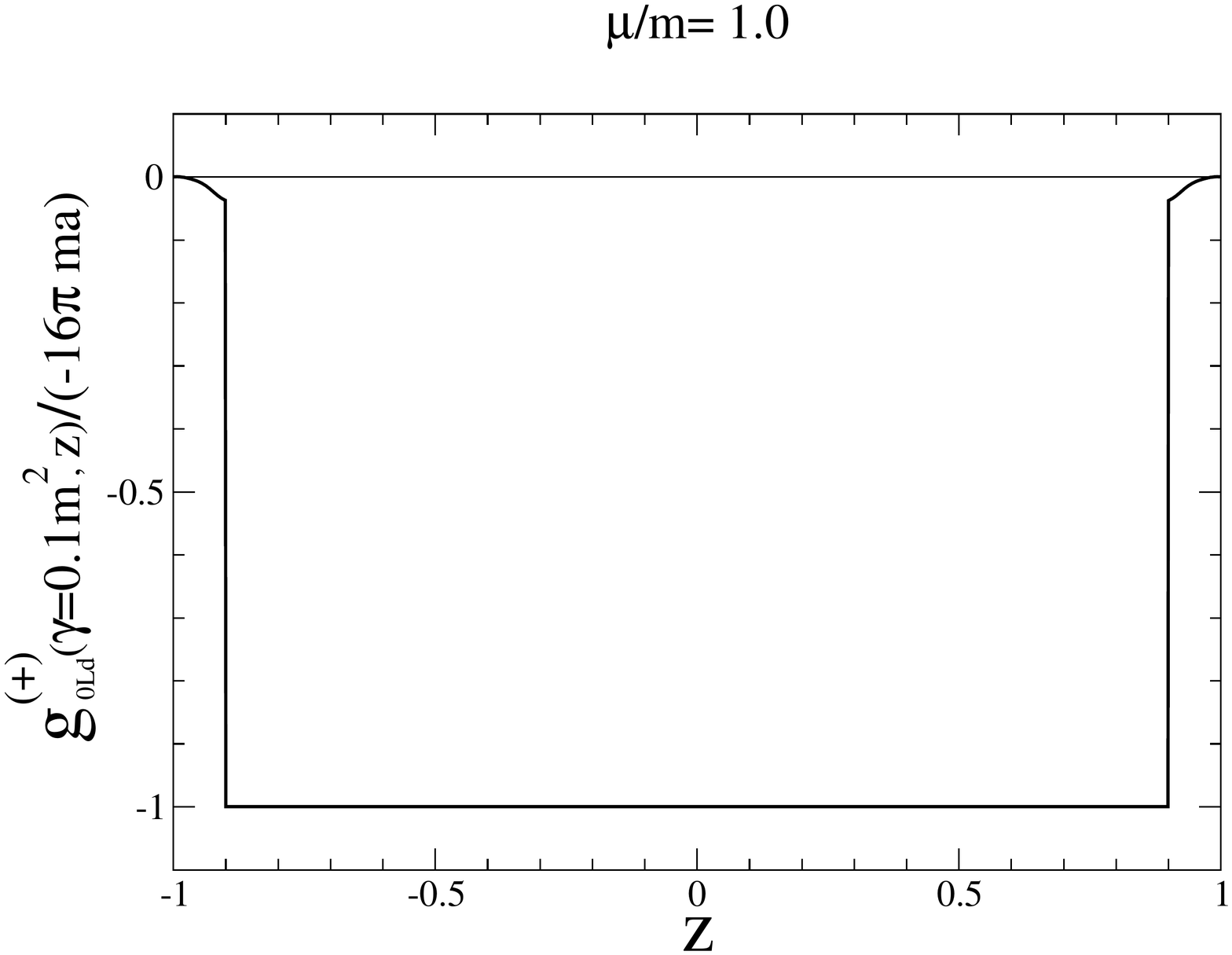}
\includegraphics[width=8.5cm] {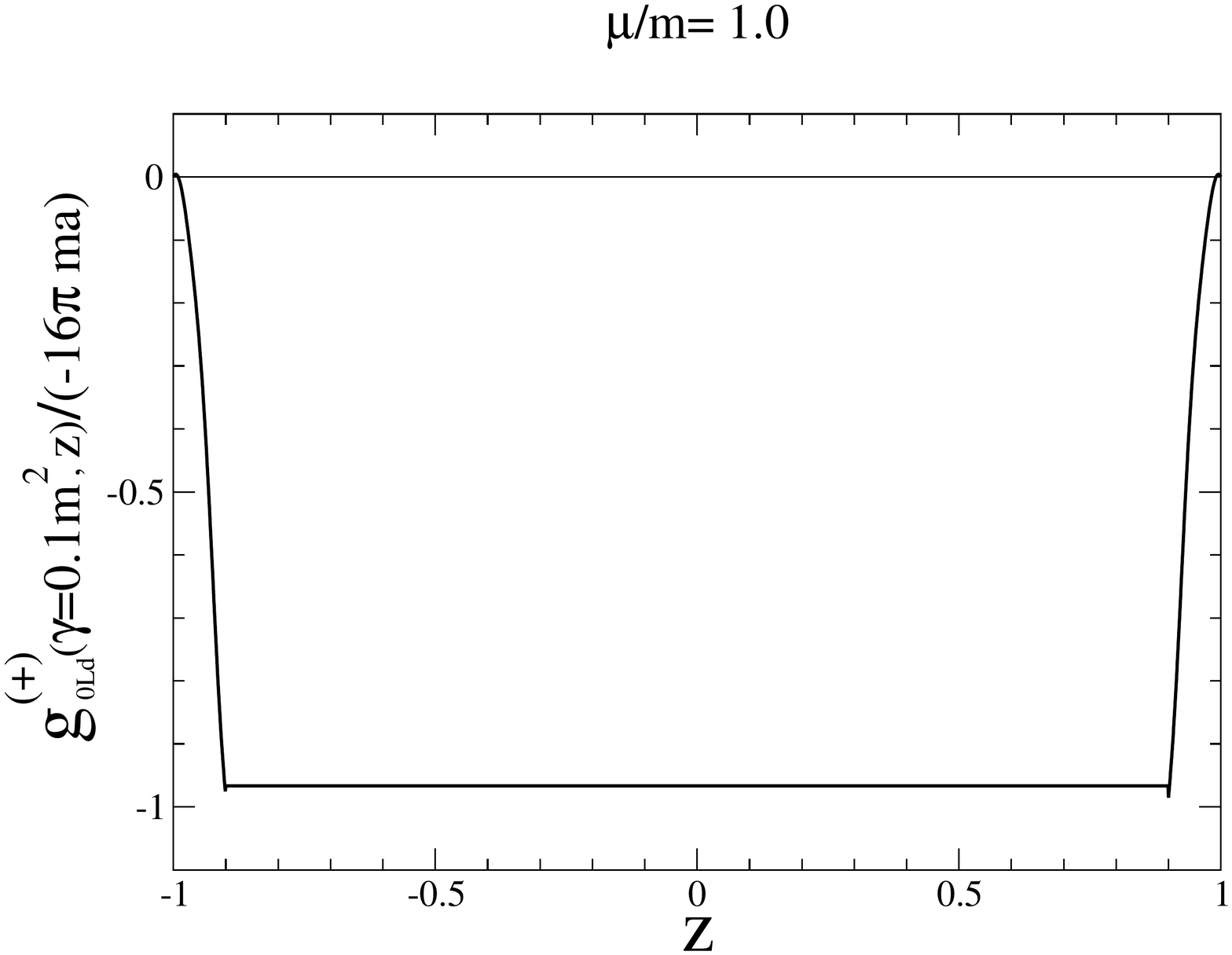}
\caption{The same as in Fig. \ref{fig1} but for running $z$, and 
$\gamma=0.1~m^2$.}
\label{fig2}\end{figure*}

\begin{figure*}[th]
\includegraphics[width=8.5cm] {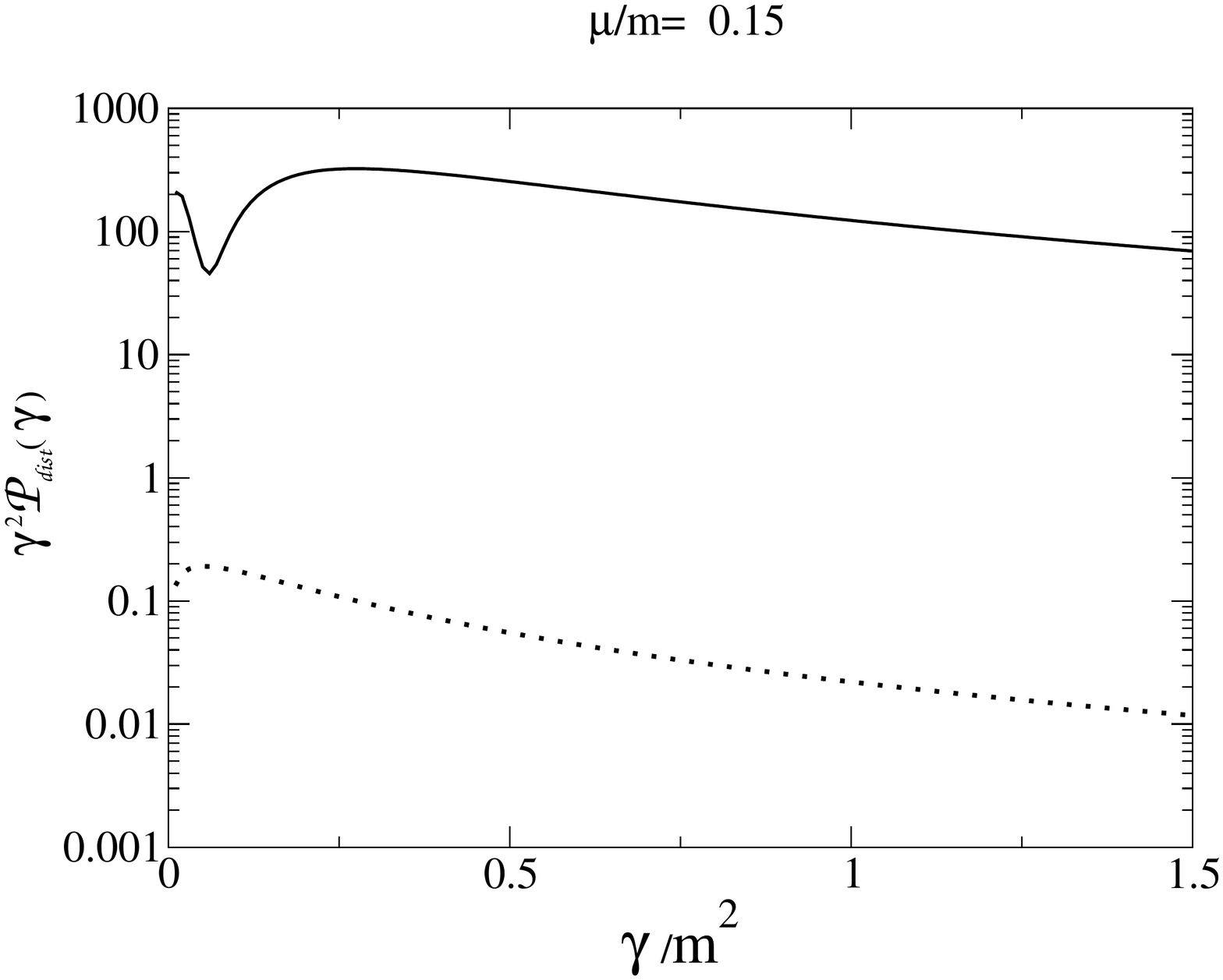}
\includegraphics[width=8.5cm] {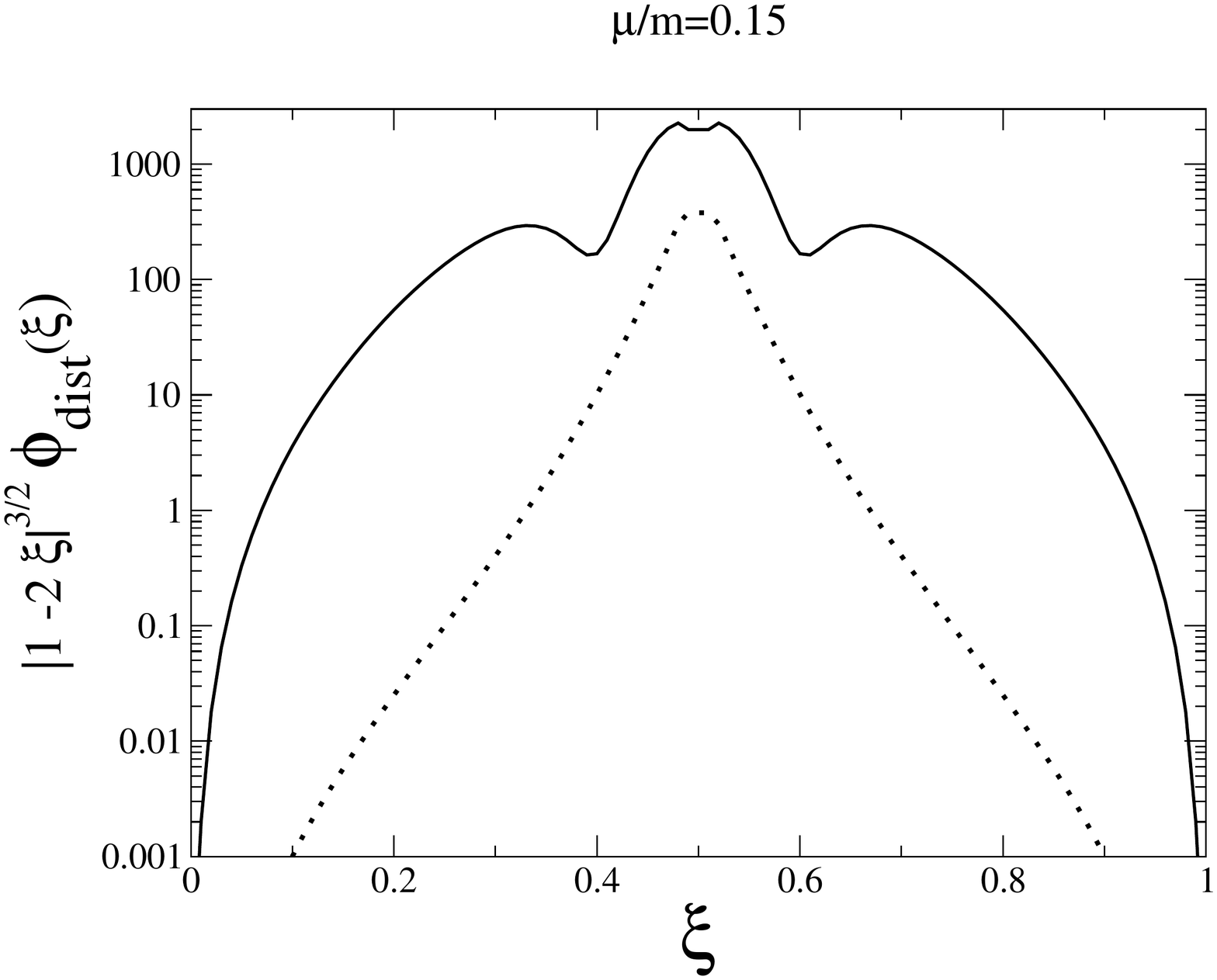}

\includegraphics[width=8.5cm] {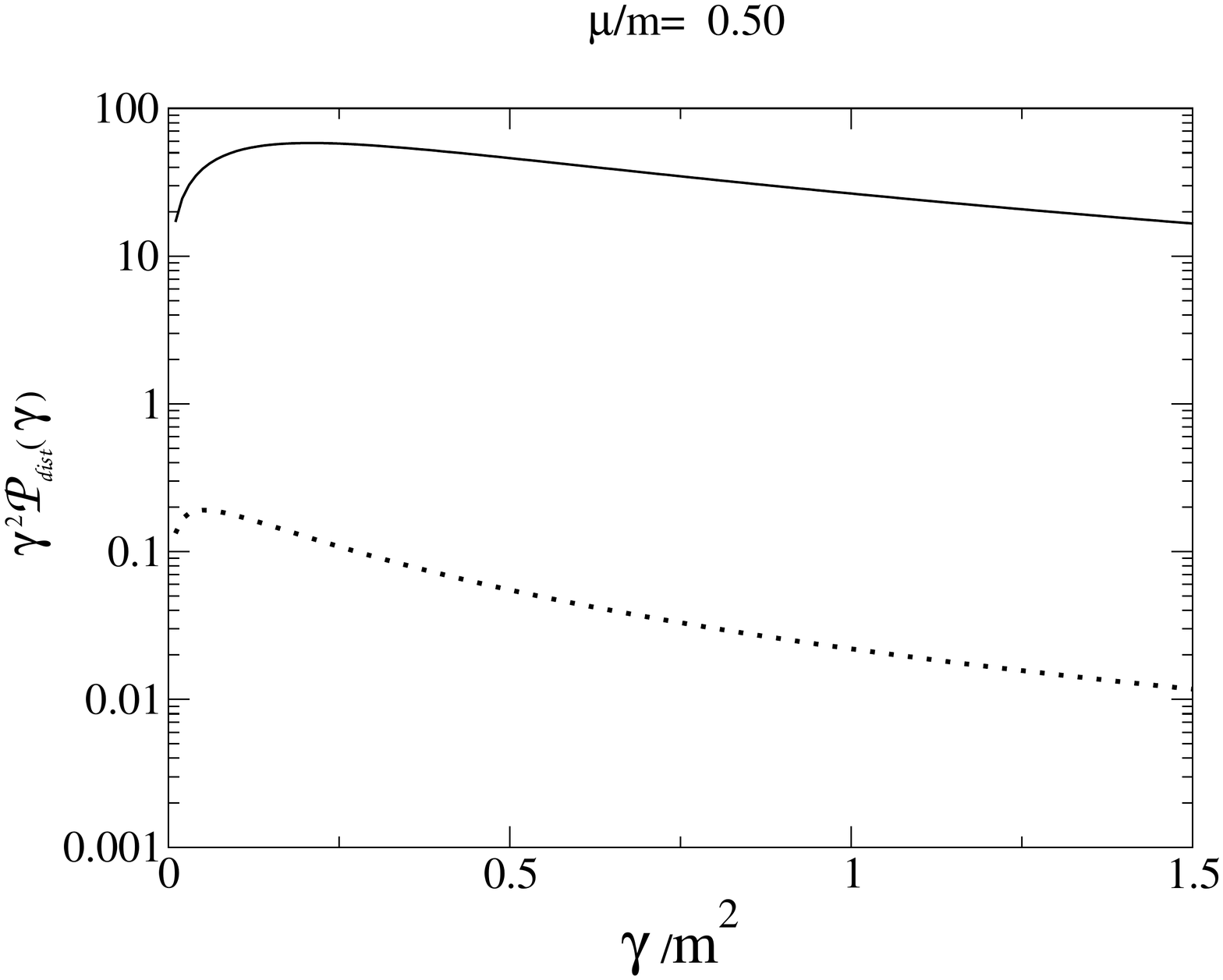}
\includegraphics[width=8.5cm] {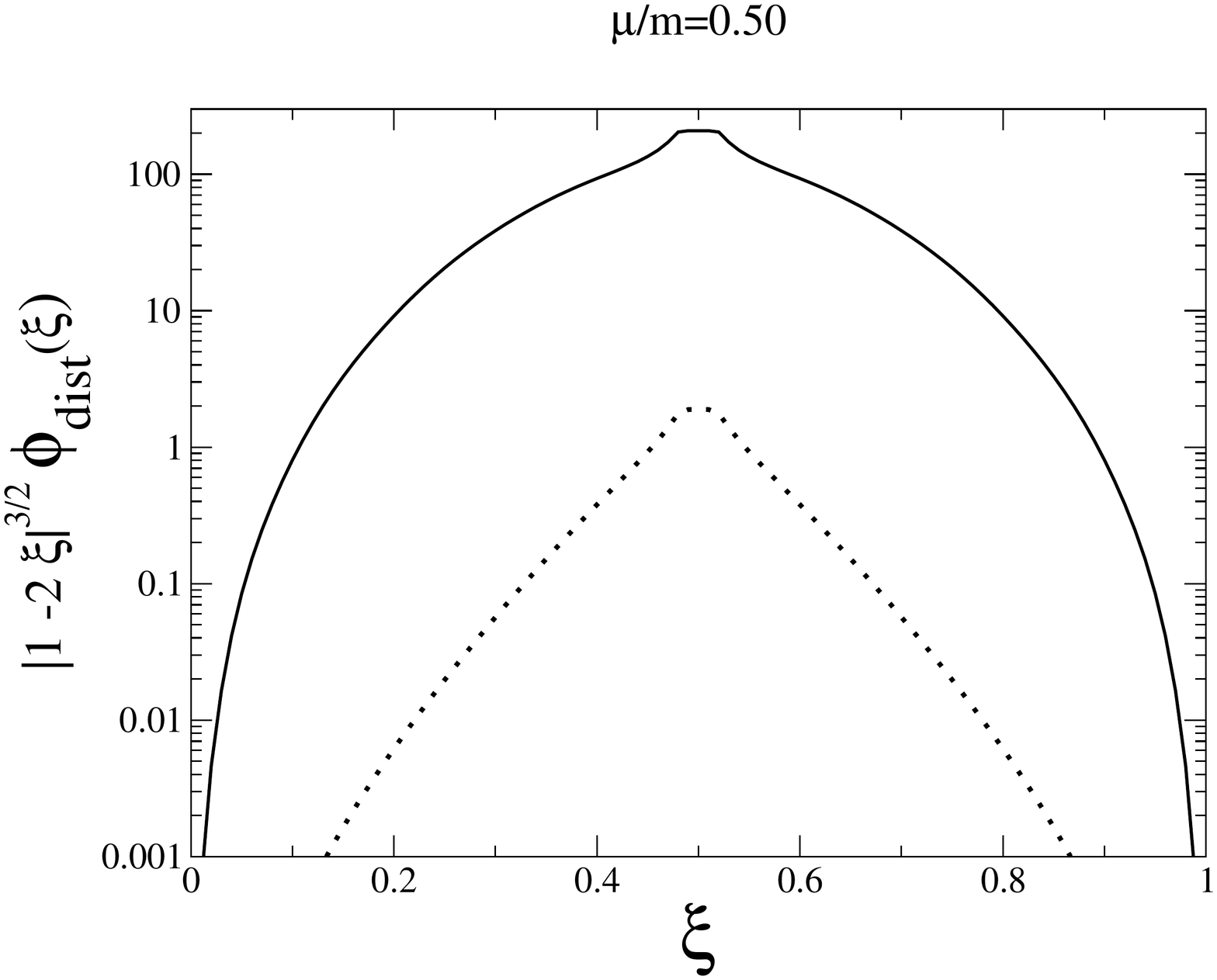}

\includegraphics[width=8.5cm] {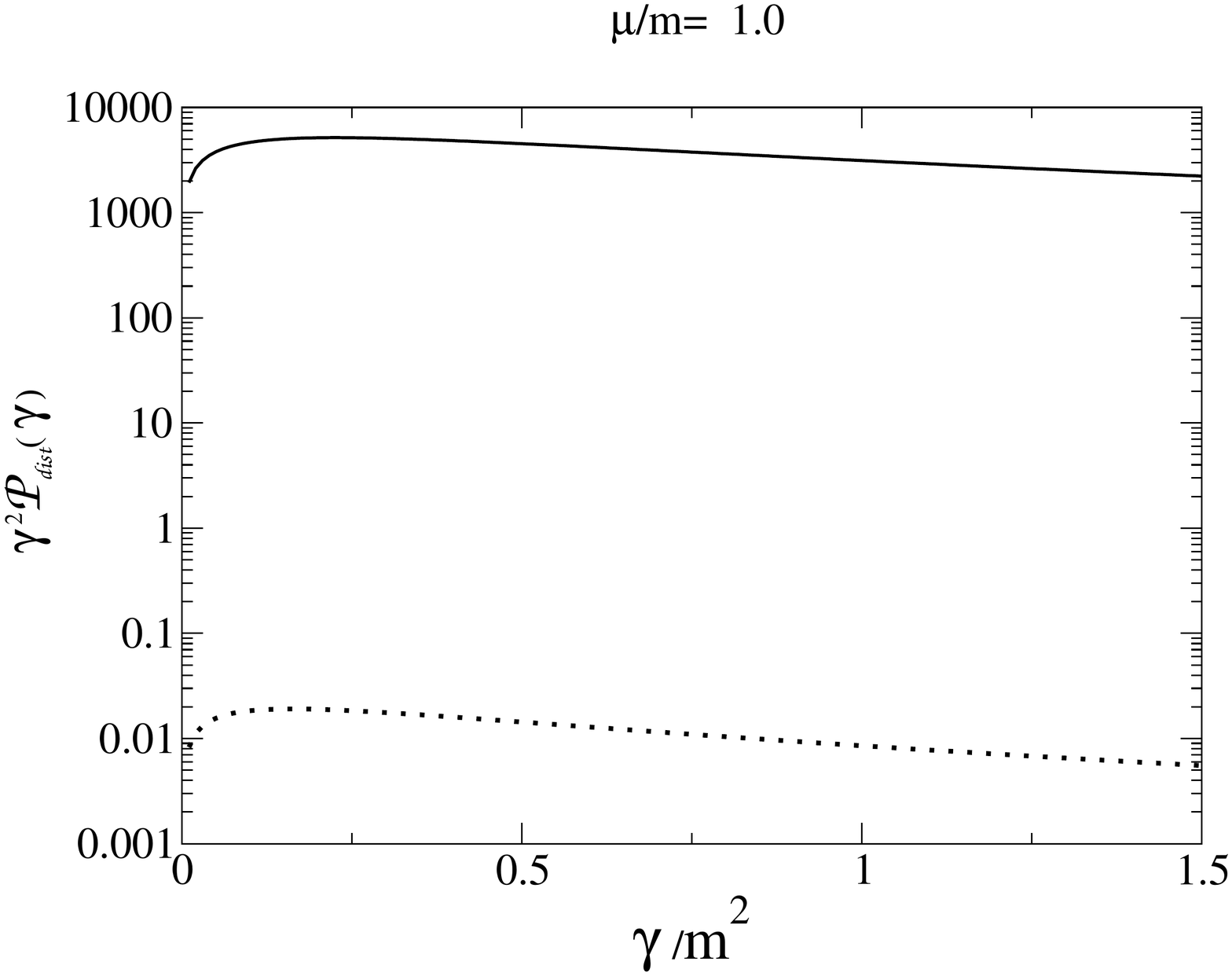}
\includegraphics[width=8.5cm] {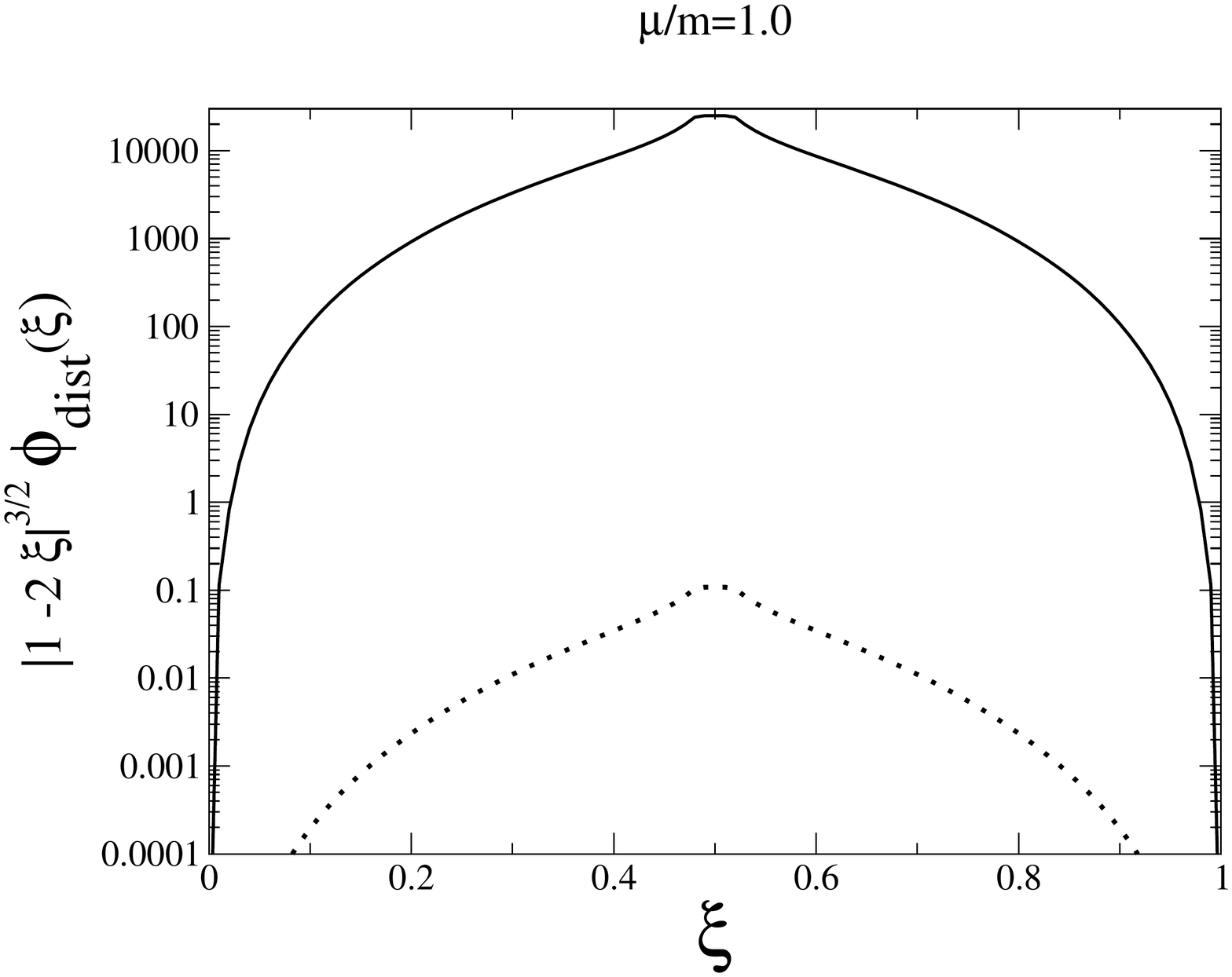}
\caption{The LF distributions obtained from the distorted part of the 3D LF
scattering wave function (see Eq. \eqref{psidis0} for
$\mu/m=0.15,~0.5,~ 1.0$ and zero-energy limit. Left panels: transverse LF-distribution $\gamma^2~{\cal P}(\gamma)$ vs the
$\gamma/m^2$ (cf the  transverse LF-distribution expression in Eq.\eqref{probgam}). Solid line: strong-interaction
regime with $\alpha=2.5$; dotted line: weak-interaction
regime with  $\alpha=0.1$
 Right panels: the same as in the left
panel, but for  $|1-2\xi|^{3/2} \phi(\xi)$, (cf the longitudinal LF-distribution in Eq. \ref{phixi})}
\label{fig3}\end{figure*}

\section{Conclusions}
\label{concl}
In the present paper,  our approach \cite{FSV1,FSV2,FSV3}, based on the Nakanishi integral representation of the Bethe-Salpeter amplitude,  is
extended for the first time to the quantitative investigation of the zero-energy limit of the inhomogeneous Bethe-Salpeter Equation, in ladder
approximation, for an interacting system composed by two massive 
scalars that exchange a massive scalar.
This achievement represents a non trivial task, that has allowed us to gain a sound confidence in the Nakanishi Ansatz, as an effective and workable tool for
obtaining actual solutions of the homogeneous and inhomogeneous BSE's in Minkowski space.
Indeed, the same approach that leads to a careful description of 
the bound states also yields
 a very accurate evaluation  of the scattering length, as shown in Tables \ref{tab1}, \ref{tab2} and
 \ref{tab3} by the quantitative comparisons  with the same observable evaluated within a totally different framework, based on the direct
calculation of the contributions from the singularities of the inhomogeneous BSE \cite{carbonell6,carbonell7}.

As in the bound state case, we have performed the calculations by using the integral representation of the BS amplitude in
terms of the Nakanishi weight function, Eq. \eqref{ptirsc}, that explicitly shows the analytic dependence of the BS amplitude
upon the invariant kinematical scalars of the scattering process, under scrutiny. Then, by applying
 the LF projection onto the null-plane to  the 
inhomogeneous BSE in Minkowski space, Eq. \eqref{inbse} (without self-energy and vertex corrections), 
 one is able
to formally obtain  the inhomogeneous
 integral equation  for the Nakanishi weight function, that
depends upon real variables. Its expression in ladder approximation is given by
 Eq. \eqref{scatlad1}. Eventually, one can
deduce another inhomogeneous integral equation for the Nakanishi weight function,  Eq. \eqref{uniq}, by  
assuming to be valid  the uniqueness
of the Nakanishi weight function, also in the non perturbative regime (recall that  the theorem was demonstrated by Nakanishi
 \cite{nak71} in a
perturbative framework, but taking into account  the whole set of infinite diagrams contributing to a given
$n$-leg amplitude).

The numerical comparisons for the scattering lengths, obtained by using  our  Eqs \eqref{zerom1} and
\eqref{gzero2}, and the corresponding quantities calculated  in Ref. \cite{carbonell6} are shown in great detail in Tables \ref{tab1}, \ref{tab2} 
and  \ref{tab3}. It has to be emphasized that the  high accuracy reached by our calculations is due to the new
decomposition  \eqref{bas1}, suitable for obtaining the numerical solutions of the two inhomogeneous integral equations,
involving the Nakanishi weight function. { The comparison with the non relativistic scattering lengths (cf Fig. \ref{a_fig})
has illustrated the potential impact of a proper treatment of the relativistic effects in the investigation of hadronic 
scattering states,
even in the low-energy regime.}

For the sake of completeness, the behavior of the Nakanishi weight functions, in the
zero-energy limit, for  weak- and strong interaction regimes have been
shown in  Figs. \ref{fig1} and \ref{fig2}. Those figures illustrate the non smooth behavior of the Nakanishi weight functions
for certain ranges of the variables, that  is an inheritance of  the singular behavior of the scattering states.
Finally, we have defined LF momentum distributions, longitudinal and transverse ones, in analogy with the bound state case,
(but without the probabilistic interpretation, entailed from the normalization of a bound state). 
Those distributions are shown in Fig. \ref{fig3}, for the sake of illustration and 
reference purpose. 
It should be noticed that the transverse LF distributions show the expected
ultraviolet behavior, i.e. a power-like one, already found in the bound state case.

In conclusion, the Nakanishi Ansatz for the BS amplitude allows one to numerically solve in a very accurate way the
inhomogeneous BSE, at least for the zero-energy limit. Such an outcome of our approach, together with the very nice results obtained for
the bound-state case, strongly encourages to move to  positive-energy scattering states, 
in order to evaluate the
phase-shifts. If the phase-shifts evaluated within our approach 
(presented elsewhere \cite{FSV4}) will agree with the ones in 
literature\cite{carbonell6}, then the
 reliability of the  Nakanishi Ansatz as a starting guess  for obtaining {\em exact solutions} of  BSE's in 
Minkowski space 
could make a substantial step forwards, confirming  the great potentiality
of this method, that can be applied to many other cases, changing dimensions \cite{Vito}, statistics, kernels, etc.

\begin{acknowledgements}
We gratefully thank  Jaume
Carbonell and Vladimir Karmanov for very stimulating discussions.
TF  acknowledges the warm hospitality of INFN Sezione di Pisa and  thanks 
 the partial financial 
support from 
 the Conselho Nacional
de Desenvolvimento Cient\'{\i}fico e Tecnol\'ogico (CNPq),
the  Funda\c c\~ao de Amparo \`a Pesquisa do
Estado de S\~ao Paulo (FAPESP). 
GS thanks the partial support of Coordena{\c c}\~ao de Aperfei{\c c}oamento  
de Pessoal de N\'\i vel Superior (CAPES) of Brazil.
MV and GS acknowledge the warm hospitality of  the  Instituto Tecnol\'ogico de Aeron\'autica, S\~ao Jos\'e dos
Campos, S\~ao Paulo, Brazil, where part of this work was performed.
\end{acknowledgements}

\newpage

\appendix
\section{Boundary properties of the Nakanishi weight function $g^{(+)}(\gamma',z',z'')$}
\label{zz'}
 This Appendix is devoted to the analysis of  the relation between (i) the Nakanishi weight function 
 $g^{(+)}(\gamma',z',z'')$, that yields the integral representation of 
 the distorted part of the 3D LF scattering wave function,
 and (ii) the weight function $\widetilde{\cal G}^+(\gamma'' ,\zeta,\zeta')$, 
 that yields the integral representation of the half-off-shell T-matrix (cf  Eq. (58) in Ref. \cite{FSV1}). Notice that the dependence upon
 $\kappa^2$ and $z_i$ has been dropped for the sake of a light notation.
 This analysis allows one to obtain the conditions fulfilled by $g^{(+)}(\gamma',z',z'')$ when $z'=\pm 1$ and $z''=\pm 1$.
 
 It is worth noting that while for $\widetilde{\cal G}^+(\gamma'' ,\zeta,\zeta')$  the constraint $\theta(1
-|\zeta|-|\zeta'|)$ holds, for the variables $z$ and $z'$ an analogous relation does not exist.

The above mentioned relation between $g^{(+)}(\gamma',z',z'')$ and $\widetilde{\cal G}^+(\gamma'' ,\zeta,\zeta')$ reads as follows
 (cf Eq. (63)
in Ref. \cite{FSV1}, where a factor of two is missing, as well as in Eq. (60), but it was not relevant for 
the formal discussion, since it can be reabsorbed in $\widetilde{\cal G}^+$)
\be
g^{(+)}(\gamma',z',z'')=
~i~2~\int_0^1 d\alpha_1 \int_0^{1}  d\alpha_2
~{1 \over ( 1-\alpha_1-\alpha_2)^3} \nonu \times\theta(1 -\alpha_1-\alpha_2)~
 \theta (1-\alpha_1-\alpha_2-|z'| -|z''-\alpha_1+\alpha_2|)
\nonu \times ~\widetilde{\cal G}^+\left({\gamma'\over (1- \alpha_1-\alpha_2)} ,{z'\over (1- \alpha_1-\alpha_2)},
{z''-\alpha_1+\alpha_2\over (1- \alpha_1-\alpha_2)}\right) \, ,
\label{gplus}\nonu\ee
where 
\be
z'=\zeta~(1-\alpha_1-\alpha_2)
\nonu
z''=\zeta'~(1-\alpha_1-\alpha_2)+\alpha_1-\alpha_2 \, .
\ee
Notice   that the constraints  $\theta (1-\alpha_1-\alpha_2)$ and $\theta(1
-|\zeta|-|\zeta'|)=\theta (1-\alpha_1-\alpha_2-|z'| -|z''-\alpha_1+\alpha_2|)$ have been explicitly written,
differently from Eq. (58) in Ref. \cite{FSV1}.

In what follows it will be shown that the above theta functions lead to a 
 vanishing   Nakanishi weight function at $|z'|=1$ or $|z''|=1$. 

Given the presence of 
$\theta (1-\alpha_1-\alpha_2-|z'| -|z''-\alpha_1+\alpha_2|)$ and $0\le
\alpha_i\le 1$, it is easily seen that for $|z'|=1$ one has
$$g^{(+)}(\gamma',z'=\pm 1,z'') = 0$$ 
The same holds for $|z''|=1$.
First of all, let us perform a change of variables, viz
\be
\xi=1-(\alpha_1+\alpha_2)~,   \quad \quad  \quad \quad\Delta=\alpha_1-\alpha_2
\nonu
\alpha_1={1-\xi+ \Delta\over 2}~, \quad \quad \quad \quad\alpha_2={1-\xi- \Delta\over 2} \, .
\ee
then Eq. \eqref{gplus} becomes
\be
\label{gplus1}
g^{(+)}(\gamma',z',z'')=~i~2~{1 \over 2}\int_0^{1} d\xi \nonu \times\int_{-1}^{1} 
 d\Delta~{\theta(1-\xi
+\Delta)~\theta(1-\xi-\Delta)
 \over \xi^3}  \nonu \times~\widetilde{\cal
G}^+\left({\gamma'\over \xi} ,{z'\over \xi},
{z''-\Delta\over \xi}\right) 
\theta (\xi-|z'| -|z''-\Delta|) \, .
\ee
For $z''=1$, one gets $z''-\Delta\geq 0$, since $\Delta\in[-1,1]$.
Then
\be\theta(1-\xi -|\Delta|)~\theta (\xi-|z'| -1\nonu+\Delta)=\theta(1-\xi -|\Delta|)~\theta [\Delta -(1 -\xi)-|z'| ]=0
\nonumber \ee
since $\Delta \geq(1 -\xi)+|z'|\geq 0$  and $1-\xi \geq |\Delta|$.
 For $z''=-1$, one has   $z''-\Delta\leq 0$, and 
then
\be\theta(1-\xi -|\Delta|)~\theta [\xi-|z'| 
\nonu-(1+\Delta)]=\theta(1-\xi -|\Delta|)~\theta [-(1-\xi)-|z'| -\Delta]=0
\nonumber \ee
since $-\Delta \geq (1-\xi)+|z'|\geq 0$  and $1-\xi \geq |\Delta|$.
Therefore, from the above results, one gets  $$g^{(+)}(\gamma',z',z''=\pm 1)=0$$.

\section{The distorted part of the 3D LF scattering wave function}
\label{distwf}
In this Appendix,  it will be shown how the expected   global free propagation 
of the constituents can be factorized out 
in the expression of $\psi^{(+)}_{dist}$, as in the non
relativistic case. This result is relevant in two respects. On one side, it
emphasizes the analogy with the non relativistic approach, and on the other side
it allows one to understand the support of the Nakanishi weight function 
$g^{(+)}(\gamma',z',z;\gamma_i,z_i)$, when $\gamma'$ runs. 

 In the CM frame ($\mbf{p}_\perp=0$ and
$p^{\pm}=M$), assuming without loss of generality  a head-on scattering, i.e. $\gamma_i=0$,
the 3D
LF scattering wave function, is given  
by \cite{FSV1}  \be
\psi^{(+)}\left(z,\gamma;\kappa^2,z_i\right)= p^+ {(1-z^2)\over
4} \int {dk^-\over 2 \pi}~ \Phi^{(+)}(k,p)=\nonu= p^+ {(1-z^2)\over
4} ~(2\pi)^3\delta^{(3)}(\tilde k-\tilde k_i) + 
\psi_{dist}(z,\gamma;\kappa^2,z_i) \, ,\nonu
\ee
where $\tilde k \equiv \{k^+, {\bf k}_\perp\}$ and $\psi_{dist}$  is
\be
\psi_{dist}(z,\gamma;\kappa^2,z_i)={(1-z^2)\over  4}
\int_{-1}^1 dz'
\int_{-\infty}^{\infty}d\gamma'\nonu \frac{g^{(+)}(\gamma',z',z;\kappa^2,z_i)}
{[\gamma'+\gamma+ z^{ 2} m^2+(1- z^{ 2})\kappa^2
 +{M^2 \over 2} z~z'  z_i  -i\epsilon]^2} \, ,
 \label{scat1}
 \nonu\ee
By using the Nakanishi weight function for the half-off-shell T-matrix, 
one gets  the following  expression \cite{FSV1}
\be
\psi_{dist}(z,\gamma;\kappa^2,z_i)=\nonu =p^+ {(1-z^2)\over
4} \int {dk^-\over 2 \pi}~\langle k^{ \mu}|G_0(p)T(p)|k^{ \mu}_i\rangle=\nonu=
~p^+ {(1-z^2)\over
4} \int {dk^-\over 2 \pi}~{i \over \left ({p \over 2} +k\right )^2-m^2+i
\epsilon}
\nonu \times{i \over \left ({p \over 2} -k\right )^2-m^2+i \epsilon}
 \int_{-1}^1 dz'\int_{-1}^1 d\zeta^\prime
\int_{-\infty}^\infty d\gamma' \nonu \times~ {\widetilde{\cal
G}^+(\gamma',z',\zeta^\prime;\kappa^2,z_i)~\theta (1 -|z'| -|\zeta'|) \over k^2+{p^2\over
4}-m^2 +\zeta^\prime p\cdot k +  z' 2 k\cdot k_i
-\gamma' +i\epsilon} \, .
\ee
Then, one can write
\be\psi_{dist}(z,\gamma;\kappa^2,z_i)=
\nonu =~-~p^+ {(1-z^2)\over
4}\int_{-1}^1 dz'\int_{-1}^1 d\zeta^\prime
\int_{-\infty}^\infty d\gamma' \nonu \times ~ \widetilde{\cal
G}^+(\gamma',z',\zeta^\prime;\kappa^2,z_i) ~
 \theta (1 -|z'| -|\zeta'|)\int {dk^-\over 2 \pi}~\nonu \times~
 {1 \over (M/2+k^+)~(M/2-k^+)}~{1\over (k^++\zeta'{M\over 2} +z' k^+_i)}\times \nonu
{1 \over(\frac{p}{2}+ k)^- - (\frac{p}{2}+ k)^-_{on}+i\epsilon/(M/2+k^+)}
\nonu \times~
{1 \over (\frac{p}{2}- k)^- - (\frac{p}{2}- k)^-_{on}+i\epsilon/(M/2-k^+)}~
\times \nonu
{1\over k^- + {k^+(\zeta'{M\over 2}+z' k^-_i)
-\gamma-\gamma' -\kappa^2 +i\epsilon\over (k^++\zeta'{M\over 2} +z' k^+_i)}} \, ,
\ee
where
\be
(\frac{p}{2}+ k)^-_{on}={2(m^2+\gamma) \over M (1-z)}
\nonu
(\frac{p}{2}- k)^-_{on}={2(m^2+\gamma) \over M (1+z)} \, .
\ee
with $k^+=-zM/2$.
Since $k^+_i=-k^-_i=-z_iM/2$, one gets
\be
\psi_{dist}(z,\gamma;\kappa^2,z_i)= 
~-~{2\over
M^2} \int_{-1}^1 dz'\int_{-1}^1 d\zeta^\prime
\int_{-\infty}^\infty d\gamma'\nonu \times  ~ \widetilde{\cal
G}^+(\gamma',z',\zeta^\prime;\kappa^2,z_i)~{\theta (1 -|z'| -|\zeta'|)
\over (\zeta' -z-z' z_i)} \nonu \times \int {dk^-\over 2 \pi}
~
{1 \over \frac{M}{2}+ k^- - (\frac{p}{2}+ k)^-_{on}+i2\epsilon/[M/(1-z)]}~~
\nonu \times {1 \over\frac{M}{2}- k^- - (\frac{p}{2}- k)^-_{on}+i2\epsilon/[M (1+z)]}~
\times \nonu
{1\over [k^- - k_{Na}^- +i2\epsilon/[M (\zeta' -z-z' z_i)]} \, ,
\ee
with
\be
k_{Na}^-={2 \over M (\zeta' -z-z' z_i)}\nonu \times ~\left[{M^2\over 4} z(\zeta'+z' z_i)
+\gamma+\gamma' +\kappa^2\right] \, .
\ee
One has the following poles (recall that $1 >z>-1$) 
\be
k_L=(\frac{p}{2}+ k)^-_{on}-{M\over 2}-i\epsilon \nonu
k_U=-(\frac{p}{2}- k)^-_{on}+{M\over 2}+i\epsilon \nonu 
k_{LU}= k_{Na}^- -i{2 \epsilon \over M (\zeta' -z-z' z_i)} \, .
\ee
In order to evaluate the analytic integration on $k^-$, one can consider the
following two cases.

If $\zeta'>z+z'z_i$,
one can close the integration contour into the upper plane, taking the residue at $k_U$, i.e.
\be\int_{-\infty}^{\infty}{d k^- \over 2\pi}~ {1 \over
\left[ k^- - k_L \right]}~ {1 \over\left[ - k^- + k_U \right]} ~ {1
\over\left [k^- -  k_{LU}\right] } 
 =\nonu=
 -i{1 \over \left[M- (\frac{p}{2}-
k)^-_{on} - (\frac{p}{2}+ k)^-_{on}+ i\epsilon\right]}\nonu \times~
 {1\over\left [ {M\over 2}-(\frac{p}{2}- k)^-_{on} - k_{Na}^- +i\epsilon\right] } =\nonu =
-i~{M^2\over 8}~(\zeta' -z-z' z_i)
\nonu \times~{(1-z^2) \over \left[\kappa^2(1-z^2)+ m^2z^2+\gamma 
-
i\epsilon\right]}~
{ (1 +z) \over  (1+\zeta' -z' z_i)}~\times \nonu {1\over 
\kappa^2(1-z^2)+ m^2z^2+\gamma   + 
 { (1 +z) \left(
  {M^2\over 2} zz'z_i
+\gamma'  \right)\over  (1+\zeta' -z' z_i)} -
i\epsilon} \, . \nonu
\ee
If $z+z'z_i>\zeta'$,
one can close the integration contour into the lower plane, taking the residue at $k_L$,
i.e.
\be
\int_{-\infty}^{\infty}{d k^- \over 2\pi}~
{1 \over
\left[ k^- - k_L \right]}~ {1 \over\left[ - k^- + k_U \right]} ~{1
\over\left [k^- -  k_{LU}\right] } 
=\nonu
=~-i {M^2\over 8}~(\zeta' -z-z' z_i)\nonu \times~{(1-z)\over (1-\zeta' +z' z_i)}~
{(1-z^2) \over \left[\kappa^2(1-z^2)+ m^2z^2+\gamma 
-
i\epsilon\right]}~ \nonu \times 
{1 \over \kappa^2(1-z^2)+m^2 z^2+\gamma  + 
{(1-z)\left({M^2\over 2} zz'
z_i
+\gamma' \right)\over (1-\zeta' +z' z_i)} ~
-i\epsilon  } \, , \nonu \ee
where $(\zeta' -z-z' z_i) \epsilon \to - \epsilon$, since $(\zeta' -z-z'
z_i)<0$.

Collecting all the above results, one gets the following expression for
$\psi_{dist}$
\be
\psi_{dist}(z,\gamma;\kappa^2,z_i)=
\nonu =~i~{(1-z^2)\over 4} ~{1\over \left[\kappa^2(1-z^2)+ m^2z^2+\gamma 
-
i\epsilon\right]}\nonu \int_{-1}^1 dz'\int_{-1}^1 d\zeta^\prime
\int_{-\infty}^\infty d\gamma' ~ \widetilde{\cal
G}^+(\gamma',z',\zeta^\prime;\kappa^2,z_i)~\times
\nonu
\Bigl[{ (1 +z) \over  (1+\zeta' -z' z_i)} \nonu \times ~
 {\theta(\zeta'-z-z'z_i  )~\theta (1 -|z'| -|\zeta'|)\over 
\kappa^2(1-z^2)+ m^2z^2+\gamma   + 
 { (1 +z) \left(
  {M^2\over 2} zz' z_i
+\gamma'  \right)\over  (1+\zeta' -z' z_i)} -
i\epsilon}
\nonu +{(1-z)\over (1-\zeta' +z' z_i)} \nonu \times ~
{\theta(z+z_iz'-\zeta')~\theta (1 -|z'| -|\zeta'|)
\over \kappa^2(1-z^2)+m^2 z^2+\gamma  + 
{(1-z)\left({M^2\over 2} zz'
z_i
+\gamma' \right)\over (1-\zeta' +z' z_i)} ~
-i\epsilon  }\Bigr] \, \nonu.
\label{psid2}\ee 
One can reobtain the expression in Eq. \eqref{scat1} by applying the Feynman 
trick to  Eq. \eqref{psid2}. For instance, one has 
\be
{\theta(\zeta'-z-z_iz' )\over \left[\kappa^2(1-z^2)+ m^2z^2+\gamma 
-
i\epsilon\right]}\nonu \times~{1\over 
\kappa^2(1-z^2)+ m^2z^2+\gamma   + 
 { (1 +z)\left(
  {M^2\over 2} zz' z_i
+\gamma'  \right) \over  (1+\zeta' -z' z_i)} -
i\epsilon}= \nonu =~\theta(\zeta'-z-z_iz' )\int_0^1 d\xi~\times
\nonu  {1\over \left[\gamma +m^2z^2+\kappa^2(1-z^2)+ \xi
{ (1 +z) \left(
  {M^2\over 2} zz' z_i
+\gamma'  \right)\over  (1+\zeta' -z' z_i)} -i \epsilon \right]^2}
\nonu ={ (1+\zeta' -z'z_i) \over (1 +z) }\int_0^1 d\alpha~ 
\theta\left[ { (1 +z) \over  (1+\zeta' -z' z_i)} -\alpha\right]
~\times \nonu
{\theta(\zeta'-z-z_i z'  )\over \left[\gamma +m^2z^2+\kappa^2(1-z^2)+ \alpha
~\left(
  {M^2\over 2} zz' z_i
+\gamma'  \right)-i \epsilon \right]^2} \, , \nonu\ee
with
$$ 1 \geq { (1 +z) \over  (1+\zeta' -z' z_i)}={ (1 +z) \over  
[1+z+(\zeta' -z-z' z_i)]}$$
since  $\theta(\zeta'-z-z_iz'  )$.
Inserting the above expression, together with the one containing $\theta(z+z_iz'-\zeta')$,
 in Eq. \eqref{psid2},
one gets the following expression for the  distorted term 
\be
\psi_{dist}(z,\gamma;z_i,\kappa^2)=\nonu =
~i~{(1-z^2)\over 4} ~\int_{-1}^1 d\zeta''\int_{-1}^1 d\zeta^\prime
\int_{-\infty}^\infty d\gamma'' ~\int_0^1 {d\alpha\over \alpha^2}
\nonu \times~ \theta(\alpha
-|\zeta''|)~\theta (1 -|{\zeta''\over \alpha}| -|\zeta'|) \nonu
\times ~{\widetilde{\cal
G}^+({\gamma'' \over \alpha} ,{\zeta''\over \alpha},\zeta^\prime)
\over\left[\gamma +m^2z^2+\kappa^2(1-z^2)+ \gamma''+
  {M^2\over 2} z\zeta'' z_i
  -i \epsilon \right]^2} \nonu
\left\{\theta\left[  (1 +z +\zeta'' z_i)  -\alpha (1+\zeta')\right]~
\theta[\alpha(\zeta'-z)-z_i\zeta''  ] + \right. \nonu \left.
\theta\left[  (1 -z -\zeta'' z_i)  -\alpha (1-\zeta')
)\right]~\theta[-\alpha(\zeta'-z)+z_i\zeta''  ]\right\}
\nonu =
~i~{(1-z^2)\over 4} ~\int_{-1}^1 d\zeta''
\int_{-\infty}^\infty d\gamma'' ~\int_0^1 {d\alpha\over \alpha^3}~ 
\int_{-1}^1 dy \nonu \times \theta (\alpha-|y|)\theta(\alpha
-|\zeta''|)~\theta (\alpha -|\zeta''| -|y|) \nonu \times ~{\widetilde{\cal
G}^+({\gamma'' \over \alpha} ,{\zeta''\over \alpha},{y \over \alpha})
\over\left[\gamma +m^2z^2+\kappa^2(1-z^2)+ \gamma''+
  {M^2\over 2} z\zeta'' z_i
  -i \epsilon \right]^2} \nonu 
\left\{\theta\left(  1 +z +\zeta'' z_i  -\alpha -y\right )~
\theta(y-\alpha z-z_i\zeta''  ) + \right. \nonu \left.
\theta\left(  1 -z -\zeta'' z_i  -\alpha +y \right]~\theta[-y+\alpha z+z_i\zeta''  )\right\}
 \, , \ee
where $\gamma''=\alpha~\gamma'$ and $\zeta''=\alpha~z'$.

The theta functions between curly brackets single out the following integration regions
\begin{itemize}
\item  $ 1 -\alpha+z +\zeta'' z_i \geq y \geq \alpha z+z_i\zeta''$ 
\item $\alpha z+z_i\zeta''\geq y \geq -(1-\alpha) +z +\zeta'' z_i$
\end{itemize}
The above intervals lead to the following constraint
$$  1 -\alpha \geq y-z -\zeta'' z_i \geq -(1-\alpha)$$
namely $\theta (1 -\alpha -| y-z -\zeta'' z_i|)$ . Then one gets
\be
\psi_{dist}(z,\gamma;z_i,\kappa^2)=
~i~{(1-z^2)\over 4} ~\int_{-1}^1 d\zeta''
\int_{-\infty}^\infty d\gamma''\nonu
\times ~\int_0^1 {d\alpha\over \alpha^3}~ 
\int_{-1}^1 dy~\widetilde{\cal G}^+({\gamma'' \over \alpha} ,{\zeta''\over \alpha},
  {y \over \alpha}) \nonu \times ~{
 \theta (\alpha -|\zeta''| -|y|)~\theta (1 -\alpha -| y-z -\zeta'' z_i|) 
\over\left[\gamma +m^2z^2+\kappa^2(1-z^2)+ \gamma''+
  {M^2\over 2} z\zeta'' z_i
  -i \epsilon \right]^2}   \, .\nonu
 \ee
The above  expression of $\psi_{dist}$ allows one
to write
the following relation between the    Nakanishi weight function 
$g^+(\gamma',z',z;\kappa^2,z_i)$, that appears in Eq. \eqref{scat1}, 
and $\widetilde{\cal G}^+$, namely  the Nakanishi weight function 
involved in the description the half-off-shell  T-matrix, 
\be
g^+(\gamma',z',z;\kappa^2,z_i)= \nonu =i\int_0^1 {d\alpha\over \alpha^3}~
 \int_{-1}^1 dy ~
 \widetilde{\cal G}^+({\gamma' \over \alpha} ,{z'\over \alpha},{y\over \alpha};\kappa^2,z_i)
 \nonu \times~\theta (\alpha -|z'| -|y|)
\theta (1 -\alpha -| y-z -z' z_i|) \, .
\label{gtildeg}\ee
 Notice that Eq. \eqref{gtildeg} can be transformed into  Eq. \eqref{gplus1} by applying a suitable change
of variables.

\section{Zero-energy limit}
\label{zerol}
The zero-energy limit of the relevant integral equations fulfilled by the
Nakanishi weight function amounts to
 consider the case  $\kappa^2=0$, namely $M^2=4m^2$. This  entails $\gamma_i=z_i=0$ through
 $M^2=4(m^2+\gamma_i)/(1-z^2_i)$. In this Appendix, the integral equations
 obtained 
 both without applying the uniqueness theorem \cite{nak71} and by exploiting it,
 are obtained following a simpler procedure than the one adopted in Ref.
 \cite{FSV1} (notice that a mistyping present in Eq. (103) of \cite{FSV1} has
 been fixed in this Appendix, as explained in what follows).
 
The Nakanishi integral equation, involving $\psi_{dist}$, for $\kappa^2 \le0$ (see \cite{FSV1})
is given by
\be 
\int_{-1}^1 dz'\int_{-\infty}^{\infty}d\gamma'' ~g^{(+)}_{(Ld)}(\gamma'',z',z;\gamma_i,z_i)
\nonu \times ~\frac{1}
{[\gamma+\gamma''+ z^{ 2}
m^2+(1- z^{ 2})\kappa^2
   +z' {M^2\over 2} z z_i-i\epsilon]^2}
=\nonu= 
g^2 ~ \int_{-1}^{1} dz'~\int^\infty_{-\infty}~d\gamma''~ \theta(-z')
~ \delta(\gamma'' -\gamma_a(z')) \nonu \times ~
{1\over
\left[\gamma+\gamma''+(1-z^2)\kappa^2 +z^2m^2 +z' {M^2\over 2} z z_i 
-i\epsilon\right]^2 } \nonu
\Bigl\{ \theta(z-z_i) ~\theta \left[1-z +z'(1-z_i)\right]
 \nonu + \theta(z_i-z)~\theta \left[1+z +z'(1+z_i)\right]\Bigr\} 
 \nonu -
{g^2 \over 2 (4 \pi)^2}~ 
\int^\infty_{-\infty}
 d\gamma'' 
\int^1_{-1}
 dz' \nonu \times{1\over \left[\gamma +\gamma''+z^2 m^2+
\kappa^2(1-z^2) +z'{M^2\over 2} z z_i
-i\epsilon   \right]^2}\nonu
\int_{-\infty}^{\infty}d\gamma'\int_{-1}^{1}d\zeta\int_{-1}^{1}d\zeta'
~g^{(+)}_{(Ld)}(\gamma',\zeta,\zeta';\kappa^2,z_i)\nonu \times
 \Bigl [{(1+z)\over (1+\zeta'-z_i\zeta)}\nonu \times
~\theta (\zeta'-z-z_i\zeta)~h'(\gamma'',z',z,z_i;\gamma',\zeta,\zeta',\mu^2)
 \nonu +{(1-z)\over (1-\zeta'+z_i\zeta)}\nonu \times 
~\theta
(z-\zeta'+z_i\zeta)~h'(\gamma'',z',-z,-z_i;\gamma',\zeta,-\zeta',\mu^2)\Bigr]
~, \nonu
\label{scatlad2}\ee 
with
\be
h'(\gamma'',z',z,z_i;\gamma',\zeta,\zeta' ,\mu^2)=~{(1+z)\over (1+\zeta'-z_i\zeta)}
~\times \nonu
\left\{{\partial \over \partial \lambda}\int ^\infty_0
{dy}~\int_0^1d\xi~\delta\left[z'-  \xi Z(z,\zeta,\zeta';z_i)\right]~\times
\right. \nonu \left. \delta\left [
{\cal F}(\lambda,y,\xi;\gamma'',z,\zeta,\zeta',\gamma';z_i,\kappa^2,\mu^2) \right]
\right \}_{\lambda=0} \, .
\ee
where
\be
{\cal F}(\lambda,y,\xi;\gamma'',z,\zeta,\zeta',\gamma';z_i,\kappa^2,\mu^2)=
\nonu=\gamma''-
\xi ~{ (1+z)\over (1+\zeta'-z_i\zeta)}
\nonu \times ~\left( {y^2 {\cal A}(\zeta,\zeta',\gamma',\kappa^2) +y(\mu^2+\gamma')
 +\mu^2 \over y}\right)
 - \xi\lambda 
\ee
For $\kappa^2=0$, it follows that $\gamma_i=z_i=0$ 
since
$$\kappa^2=m^2 - {M^2\over 4}=\left({p\over 2} \pm k_i\right)^2 - {M^2\over 4}=
-\gamma_i-z^2_i {M^2\over 4}=0$$

Then, taking into account that 
$g^{(+)}_{(Ld)}(\gamma,z',z;\kappa^2=0)$  has support only for positive 
$\gamma$, one can 
write (cf Eq. \eqref{scatlad2} and subsec. \ref{supp})
\be
\int_{-1}^1 dz'\int_{0}^{\infty}d\gamma''
~~\frac{g^{(+)}_{(Ld)}(\gamma'',z',z;\kappa^2=0)}
{[\gamma+\gamma''+ z^{ 2}
m^2 -i\epsilon]^2}
=\nonu= 
g^2 ~ \int^\infty_{0}~d\gamma''~ \int_{-1}^{1} dz'~{\theta(-z')
~ \delta(\gamma'' +z' \mu^2)
\over
\left[\gamma+\gamma'' +z^2m^2   -i\epsilon\right]^2 } \nonu \times~
\Bigl[ \theta(z) ~\theta \left(1-z +z'\right)
 + \theta(-z)~\theta \left(1+z +z'\right)\Bigr] \nonu -
{g^2 \over 2(4\pi)^2}~
\int_{0}^{\infty}d\gamma'\int_{-1}^{1}d\zeta \nonu \times \int_{-1}^{1}d\zeta'
~g^{(+)}_{(Ld)}(\gamma',\zeta,\zeta';\kappa^2=0)
\nonu \times\int^\infty_{-\infty}
 d\gamma''
~{1\over \left[\gamma +\gamma''+z^2 m^2 
-i\epsilon   \right]^2}\int^1_{-1}
 dz' \nonu \times ~\Bigl [{(1+z)\over (1+\zeta')}
~\theta (\zeta'-z)~{\cal Z}'(\gamma'',z',z;\gamma',\zeta',\mu^2)
\nonu+{(1-z)\over (1-\zeta')}
~\theta
(z-\zeta')~{\cal Z}'(\gamma'',z',-z;\gamma',-\zeta',\mu^2)\Bigr] \, ,
 \nonu\label{zero1}\ee
where
 the kernel ${\cal Z}'$ is given by
\be
{\cal Z}'(\gamma'',z',z;\gamma',\zeta',\mu^2)= {(1+z)\over
(1+\zeta')}~\times \nonu\Bigl\{
{\partial \over \partial \lambda} 
\int_0^\infty {d y}~\int_0^1 d\xi \,
\delta \left[\gamma'' - \xi \Gamma_0(y,z,\zeta',\gamma')
-\xi \lambda\right]\nonu
\times\delta \left[z'-\xi { (1+z)\over (1+\zeta')}~\zeta
\right]    \Bigr\}_{\lambda=0}
\label{zcal}\ee
with
\be
\Gamma_0(y,z,\zeta',\gamma')= { (1+z)\over (1+\zeta')}~{1 \over y}
\nonu \times
\left\{y^2 ~{\cal A}_0(\zeta',\gamma' )+y(\mu^2
+\gamma') +\mu^2\right\}
\ee
and \be
{\cal A}_0(\zeta',\gamma' )={\zeta'}^{2} \frac{M^2}{4} +
\gamma'=~{\zeta'}^{2} m^2 +
\gamma'~\ge~0 \, ,
\label{cala}\ee
Notice that $\gamma'$  is positive  and therefore also
 $\Gamma_0$  has to be positive. Finally, $\gamma''$ in Eq. \eqref{zcal} has to
 be positive 
for getting a non vanishing ${\cal Z}'(\gamma'',z',z;\gamma',\zeta',\mu^2)$.

Performing (i) the integration on $z'$ in both sides of Eq. \eqref{zero1} 
(recall that $1 >|\xi \zeta
(1\pm z)/(1\pm \zeta')|$) and (ii)  the
integration on $\zeta$ in the rhs, one gets
\be
\int_{0}^{\infty}d\gamma''
\frac{g^{(+)}_{0Ld}(\gamma'',z)}
{[\gamma+\gamma''+ z^{ 2}
m^2 -i\epsilon]^2}
=\nonu= 
{g^2\over \mu^2} ~\int^\infty_{-\infty}~d\gamma''~
{\theta(\gamma'')
\over
\left[\gamma+\gamma'' +z^2m^2   -i\epsilon\right]^2 } \nonu \times~
\Bigl \{ \theta(z) ~\theta \left[1-z -\gamma''/\mu^2\right]
 \nonu + \theta(-z)~\theta \left[1+z  -\gamma''/\mu^2\right]\Bigr\} +\nonu -
{g^2 \over 2(4\pi)^2}~
\int_{0}^{\infty}d\gamma'\int_{-1}^{1}d\zeta'~g^{(+)}_{0Ld}(\gamma',\zeta')
\nonu \times~\int^\infty_{-\infty}
 d\gamma''~\theta(\gamma'')~
~{1\over \left[\gamma +\gamma''+z^2 m^2 
-i\epsilon   \right]^2}
 \times \nonu\Bigl [{(1+z)\over (1+\zeta')}
~\theta (\zeta'-z)~{h}_0'(\gamma'',z;\gamma',\zeta',\mu^2)
 \nonu+{(1-z)\over (1-\zeta')}
~\theta
(z-\zeta')~{h}_0'(\gamma'',-z;\gamma',-\zeta',\mu^2)\Bigr] \, ,
\label{zero2}\ee
where
\begin{enumerate}
\item\be
g^{(+)}_{0Ld}(\gamma'',z)=\int_{-1}^1 dz'
g^{(+)}_{(Ld)}(\gamma'',z',z;\kappa^2=0) \, .
\ee
\item
\be
\int_{-1}^{1} dz'~\theta(-z')
~ \delta(\gamma'' +z' \mu^2)=\nonu = {1 \over \mu^2}
\theta(\gamma'')\theta(\mu^2-\gamma'')  \, .\ee
\item 
\be
{h}_0'(\gamma'',z;\gamma',\zeta',\mu^2)=
{(1+z)\over
(1+\zeta')}~\left .
{\partial \over \partial \lambda} 
\int_0^\infty {d y}\right. \nonu \left.~\int_0^1 d\xi \,
\delta \left[\gamma'' - \xi \Gamma_0(y,z,\zeta',\gamma')
-\xi \lambda\right]\,   \right|_{\lambda=0} \, .\nonu\ee
\end{enumerate}
From Ref. \cite{FSV2}, one recognizes that 
${h}_0'(\gamma'',z;\gamma',\zeta',\mu^2)$ is
the suitable kernel for a bound state with zero energy. Therefore
one can write 
 \be
h^\prime_0(\gamma'',z;\gamma',\zeta',\mu^2) = \nonu =
\theta \left[-~{\cal B}_0(z,\zeta',\gamma',\gamma'',\mu^2) 
 -2 \mu \sqrt{ {\zeta'}^{2} \frac{M^2}{4} +
\gamma'} \right]
 \nonu \times
\Bigl[- {{\cal B}_0(z,\zeta',\gamma',\gamma'',\mu^2) \over
{\cal A}_0(\zeta',\gamma')~\Delta_0(z,\zeta',\gamma', \gamma'',
\mu^2 ) }~{1\over\gamma''}  + { (1+\zeta')\over (1+z)} \nonu \times ~
 \int_{y_-}^{y_+}{dy}~{y^2 \over \left[ {y}^2{\cal A}_0(\zeta',\gamma' )+
  y(\mu^2 +\gamma')+\mu^2
\right ]^2}\Bigr]\nonu
-{ (1+\zeta')\over (1+z)}~
\int_0^{\infty}{dy}~{y^2 \over \left[ {y}^2{\cal A}_0(\zeta',\gamma' )+ y(\mu^2 +\gamma')+\mu^2
\right ]^2} \, ,\nonu 
\label{gzero3}\ee
with (see also Eq. \eqref{cala})
\be
{\cal B}_0(z,\zeta',\gamma', \gamma'',\mu^2 )=
\mu^2+\gamma' -\gamma''
{ (1+\zeta')\over (1+z)}~\leq 0~~,
\nonu
\Delta^2_0(z,\zeta',\gamma', \gamma'',\mu^2 )=\nonu =
{\cal B}_0^2(z,\zeta',\gamma',\gamma'',\mu^2)
- 4 \mu^2~ {\cal A}_0(\zeta',\gamma')~\geq ~0 ~~,
\nonu
y_\pm=
{1 \over 2{\cal A}_0(\zeta',\gamma')} \nonu \times ~
 \left[ -{\cal B}_0(z,\zeta',\gamma',\gamma'',\mu^2)
 \pm \Delta_0(z,\zeta',\gamma', \gamma'',\mu^2 )\right] \, .
\label{gzero4}\ee

Notice that in the inhomogeneous term in Eq. \eqref{zero2} the factor
$\theta(\mu^2-\gamma'')$ has been dropped, given the presence of the step functions $\theta\left[1\pm z  -\gamma''/\mu^2\right]$. In Eq. (103)
of Ref. \cite{FSV1} the step function $\theta(\mu^2-\gamma'')$ has been
accidentally overlooked.

In conclusion, by applying the Nakanishi theorem   on the uniqueness of the
weight function \cite{nak71}, one has the following integral equation
\be
g^{(+)}_{0Ld}(\gamma,z)
=
{g^2\over \mu^2} ~
\theta(\gamma)
~
\theta \left[\mu^2(1-|z|) -\gamma\right]
  \nonu -
{g^2 \over 2(4\pi)^2}~\theta(\gamma)~
\int_{0}^{\infty}d\gamma'\int_{-1}^{1}dz'~g^{(+)}_{0Ld}(\gamma',z')
~
 \times \nonu\Bigl [{(1+z)\over (1+z')}
~\theta (z'-z)~{h}_0'(\gamma,z;\gamma',z',\mu^2)
\nonu  +{(1-z)\over (1-z')}
~\theta
(z-z')~{h}_0'(\gamma,-z;\gamma',-z',\mu^2)\Bigr] \,.
\label{zero3}
\ee
\section{An effective decomposition of $g^{(+)}_{0Ld}(\gamma,z)$}
\label{mdeco}
In this Appendix, the decomposition of $g^{(+)}_{0Ld}(\gamma,z)$ shown in Eq.
\eqref{bas1} is applied to the simple case of  Eq. \eqref{gzero2}, based on the Nakanishi
uniqueness theorem \cite{nak71}, in order to give the explicit
representation of the numerical system to be solved.

Inserting the decomposition \eqref{bas1}, 
\be
g^{(+)}_{0Ld}(\gamma,z)=~\beta~\theta(-t)+
\theta(t)~\sum_{\ell=0}^{N_z} \sum_{j=0}^{N_g}~
A_{\ell j} ~G_\ell(z) ~{\cal L}_j(t) \, ,\nonu
\label{mdeco1}\ee
with $t=\gamma -\mu^2(1-|z|)$, in Eq. \eqref{gzero2}, given by  (notice that in the following expression, the symmetry properties of both the weight
function and the kernel ${h}_0'$ are exploited),
\be
g^{(+)}_{0Ld}(\gamma,z)
= 
{g^2\over \mu^2} ~
\theta(\gamma)
~
\theta \left[\mu^2(1-|z|) -\gamma\right]
  \nonu -
{g^2 \over (4\pi)^2}~\theta(\gamma)~
\int_{0}^{\infty}d\gamma'\int_{-1}^{1}dz'~g^{(+)}_{0Ld}(\gamma',z')
~
 \times \nonu {(1+z)\over (1+z')}
~\theta (z'-z)~{h}_0'(\gamma,z;\gamma',z',\mu^2) \, ,
\ee
one can quickly obtain
the following coupled system
\be
\beta ~\theta(-t)={g^2\over \mu^2} ~
\theta(\gamma)
~\theta(-t)
   -
{g^2 \over (4\pi)^2}~\theta(\gamma)~\theta(-t)~\beta \nonu \times ~
\int_{-1}^{1}dz' \int_{0}^{ \mu^2 (1-|z'|)}d\gamma'~
 \nonu \times ~{(1+z)\over (1+z')}
~\theta (z'-z)~{h}_0'(\gamma,z;\gamma',z',\mu^2)  
\nonu
 -
{g^2 \over (4\pi)^2}~\theta(\gamma)~\theta(-t) ~
\sum_{\ell=0}^{N_z} \sum_{j=0}^{N_g}~
A_{\ell j} ~
\int_{-1}^{1}dz'
\nonu \times ~\int_{\mu^2 (1-|z'|)}^{\infty}d\gamma'~
G_\ell(z') ~{\cal L}_j\left[\gamma'-\mu^2 (1-|z'|)\right]
~
 \times \nonu{(1+z)\over (1+z')}
~\theta (z'-z)~{h}_0'(\gamma,z;\gamma',z',\mu^2) \,,
\label{beta}\ee
and
\be
\theta(t)~\sum_{\ell=0}^{N_z} \sum_{j=0}^{N_g}~
A_{\ell j} ~G_\ell(z) ~{\cal L}_j(t)
=\nonu = -
{g^2 \over (4\pi)^2}~\theta(\gamma)~\theta(t)~\beta
\int_{-1}^{1}dz' \int_{0}^{ \mu^2 (1-|z'|)}d\gamma'~
 \times \nonu{(1+z)\over (1+z')}
~\theta (z'-z)~{h}_0'(\gamma,z;\gamma',z',\mu^2)  +
\nonu
 -
{g^2 \over (4\pi)^2}~\theta(\gamma)~\theta(t) ~
\sum_{\ell=0}^{N_z} \sum_{j=0}^{N_g}~
A_{\ell j} ~
\int_{-1}^{1}dz' \nonu\int_{\mu^2 (1-|z'|)}^{\infty}d\gamma'~
G_\ell(z') ~{\cal L}_j\left[\gamma'-\mu^2 (1-|z'|)\right]
~
 \times \nonu{(1+z)\over (1+z')}
~\theta (z'-z)~{h}_0'(\gamma,z;\gamma',z',\mu^2) \, .
\label{aij}\ee

For $\gamma=0$ and $z=0$, Eq. \eqref{beta} reduces to
\be
\beta ={g^2\over \mu^2} ~
   +\beta~ I_{\beta,\beta} 
 +\sum_{\ell=0}^{N_z} \sum_{j=0}^{N_g}~
I_{\beta,\ell j}~A_{\ell j}  \, ,
\label{beta1}
\ee
where 
\be
I_{\beta\beta}=~-
{g^2 \over (4\pi)^2}~\int_{0}^{1}dz' \int_{0}^{ \mu^2 (1-z')}d\gamma'
\nonu \times ~
 {1\over (1+z')}
~~{h}_0'(\gamma=0,z=0;\gamma',z',\mu^2) 
\ee
and 
\be
I_{\beta,\ell j}=~-
{g^2 \over (4\pi)^2}\nonu \times~\int_{0}^{1}dz'\int_{\mu^2 (1-z')}^{\infty}d\gamma'~
G_\ell(z')  ~{\cal L}_j\left[\gamma'-\mu^2 (1-z')\right] \nonu \times 
~{1\over (1+z')}
~{h}_0'(\gamma=0,z=0;\gamma',z',\mu^2) \, ,
\ee
with (cf Eq. \eqref{gzerom})
\be
{h}_0'(\gamma=0,z=0;\gamma',z',\mu^2)=-{ (1+z')}\nonu \times ~
\int ^\infty_0 {dy}~{ y^2\over \left[y^2(\gamma' +m^2z'^2) +y(\gamma'+\mu^2)
+\mu^2\right]^2} \, .\nonu
\ee
A matrix representation can be obtained for Eq. \eqref{aij} by multiplying both
sides by $ G_{\ell '}(z) ~{\cal L}_{j'}(t)$ and integrating, viz
\be
A_{\ell ' j'} 
= ~\beta~ I_{\ell' j',\beta}+ 
\sum_{\ell=0}^{N_z} \sum_{j=0}^{N_g}~
I_{\ell' j',\ell j}~A_{\ell j} 
\ee
where (cf Eq. \eqref{gzerom})
\be
I_{\ell' j',\beta}=~-
{g^2 \over (4\pi)^2}~\int_{-1}^{1}dz \int_{0}^\infty dt~G_{\ell '}(z) ~{\cal L}_{j'}(t)
\int_{z}^{1}dz' \nonu\int_{0}^{ \mu^2 (1-|z'|)}d\gamma'~
 {(1+z)\over (1+z')}
~{h}_0'(\gamma,z;\gamma',z',\mu^2)
\ee
and 
\be
I_{\ell' j',\ell j}=~-
{g^2 \over (4\pi)^2}~\int_{-1}^{1}dz \int_{0}^\infty dt~G_{\ell '}(z) ~{\cal L}_{j'}(t)
\nonu \int_{z}^{1}dz' \int_{\mu^2 (1-|z'|)}^{\infty}d\gamma'~
G_\ell(z') ~\times \nonu {\cal L}_j\left[\gamma'-\mu^2 (1-|z'|)\right]
~{(1+z)\over (1+z')}
~{h}_0'(\gamma,z;\gamma',z',\mu^2) \, . \nonu
\ee

\end{document}